\RequirePackage{fix-cm}
\documentclass{svjour3}                     
\smartqed  
\usepackage{natbib}
\usepackage{multirow}
\usepackage{graphicx}
\usepackage{amsmath}

\usepackage{xcolor}
\usepackage{ulem} 

\newcommand{\hl}[1]{#1}

\makeatletter
\renewcommand*\env@matrix[1][\arraystretch]{%
  \edef\arraystretch{#1}%
  \hskip -\arraycolsep
  \let\@ifnextchar\new@ifnextchar
  \array{*\c@MaxMatrixCols c}}
\makeatother
\journalname{Rock Mechanics and Rock Engineering}

\begin{document}

\title{Inducing tensile failure of \hl{claystone} through thermal pressurization in a novel triaxial device}

\titlerunning{Inducing tensile failure of claystone through thermal pressurization in a novel triaxial device}


\author{Philipp Braun         \and
        Pierre Delage         \and
        Siavash Ghabezloo     \and
        Baptiste Chabot		\and
        Nathalie Conil   \and
        Minh-Ngoc Vu   
}


\institute{P. Braun         \and
        	S. Ghabezloo     \and
        	P. Delage        \and 
        	B. Chabot        \and 
            \at  
            Laboratoire Navier, 6-8 avenue Blaise-Pascal, Cité Descartes 77455 Champs-sur-Marne, Paris, France\\
             \email{philipp.braun@enpc.fr}          
            \and
            P. Braun \and 
			M.-N. Vu        
			\at
              Andra, 1 Rue Jean Monnet, 92290 Châtenay-Malabry, France 
              \and
              N. Conil
              \at
              Andra, Centre de Meuse/Haute-Marne BP9, 55290 Bure, France
              \at
               \emph{Now at:} INERIS, Ecole des Mines, Parc de Saurupt, 54042 Nancy, France
}

\date{Accepted manuscript DOI: 10.1007/s00603-022-02838-3}

\maketitle

\section*{Highlights}
\begin{itemize}
\item Thermal pressurization of clay rock in deep radioactive waste repositories can reduce the effective stresses, which can lead to damage or failure.
\item Our novel laboratory triaxial device is able mimic in situ conditions: Constant vertical total stress, zero lateral deformation and thermal pressurization. 
\item Pore pressure increase, vertical extension strains and thermal pressurization failure were recorded in a series of tests on Callovo-Oxfordian claystone specimens.
\item The effective tensile \hl{strength} was reached at values around 3 MPa in tension and temperatures between 53 and 64 °C, creating sub-horizontal fractures. 
\item The experimental responses can be well reproduced using a thermo-poroelasticity model.
\item Hoek-Brown and Fairhurst generalized Griffith criteria appear suitable to account for the rock’s tensile resistance.
\end{itemize}

\clearpage

\begin{abstract}
Complex coupled thermo-hydromechanical (THM) loading paths are expected to \hl{occur in} clay rocks \hl{which serve} as host formations for geological radioactive waste repositories. Exothermic waste packages heat the rock, causing thermal strains and temperature induced pore pressure \hl{build-up}. The drifts are designed \hl{in such a way} as to limit these effects. 
One has to anticipate failure and fracturing of the material, should pore pressures exceed the tensile resistance of the rock. To characterise the behaviour of the Callovo-Oxfordian claystone (COx) under effective tension and to quantify the tensile failure criterion, a laboratory program is carried out in this work. THM loading paths \hl{which} correspond to the expected in situ conditions are \hl{recreated in the laboratory}. To this end, a special triaxial system was developed, which \hl{allows the} independent control of radial and axial stresses, as well as \hl{of} pore pressure and temperature of rock specimens. More importantly, the device allows \hl{one} to maintain axial effective tension on a specimen. Saturated cylindrical claystone specimens were tested in undrained conditions under constrained lateral deformation and under nearly constant axial stress. The specimens were heated until the induced pore pressures created effective tensile stresses and ultimately fractured the material. \hl{The failure happened at average} axial effective tensile stresses around 3.0 MPa. Fracturing under different lateral total stresses allow\hl{s one} to describe the failure with a Hoek-Brown or Fairhurst's generalized Griffith criterion. Measured axial extension strains are analysed based on a transversely isotropic thermo-poroelastic constitutive model, which is able to satisfactorily reproduce the observed behaviour.  

\end{abstract}

\clearpage
\section*{\hl{List of Symbols}}

\hl{Note that the matrix notation is used throughout this work.}\\
\begin{tabular}{ ll }
$h$	&	Direction parallel to bedding 					\\
$z$	&	Direction perpendicular to bedding					\\
$\sigma_{i}$	&	Total stress in direction $i$					\\
$\sigma'_{i}$	&	Terzaghi effective stress in direction $i$					\\
$\varepsilon_i$	&	Strain in direction $i$					\\
$\sigma_c$	&	Unconfined compressive strength					\\
$\sigma_t$	&	Tensile strength					\\
$E_i$	&	Young's modulus in direction $i$					\\
$\nu_i$	&	Poisson's ratio in direction $i$					\\
$m_i$	&	Hoek-Brown criterion parameter					\\
$\phi$	&	Porosity					\\
$\rho$	&	Wet density					\\
$\rho_d$	&	Dry density					\\
$w$	&	Water content					\\
$S_r$	&	Degree of saturation					\\
$s$	&	Suction					\\
$p_f$	&	Pore pressure					\\
$T$	&	Temperature					\\
$\sigma$	&	Isotropic confining  pressure					\\
$\varepsilon_{hyd}$	&	Volumetric hydration swelling					\\
$\gamma_t$	&	Fracture angle with respect to bedding					\\
$C_{ij}$	&	Elastic compliance matrix					\\
$b_i$	&	Biot's coeffcient in direction $i$					\\
$\alpha_{d,i}$	&	Drained thermal expansion coeffcient in direction $i$					\\
$G'$	&	Shear modulus within the isotropic plane					\\
$G$	&	Shear modulus perpendicular to the isotropic plane					\\
$\alpha_{\phi}$	&	Bulk thermal expansion coefficient of the pore volume					\\
$M$	&	Biot's modulus					\\
$K_f$	&	Bulk modulus of the pore fluid					\\
$K_{\phi}$	&	Bulk modulus of the pore volume					\\
$\varphi$	&	Friction angle					\\
$c$	&	Cohesion					\\
$V_L$	&	Volume of the drainage system					\\
$c_L$	&	Compressibility of the drainage system with respect to fluid pressure					\\
$p_L$	&	Fluid pressure within the drainage system					\\
$\kappa_L$	&	Compressibility of the drainage system with respect to radial confining pressure					\\
$\sigma_{rad}$	&	Radial confining pressure					\\
$\alpha_L$	&	Bulk thermal expansion coefficient of the drainage system					\\
$M_s$	&	Fluid mass within the specimen					\\
$m_f$	&	Fluid mass per unit volume of the specimen					\\
$V_s$	&	Total specimen volume					\\
$M_L$	&	Fluid mass within the drainage system					\\
$\rho_L$	&	Fluid density within the drainage system					\\
$\rho_f$	&	Pore fluid density					\\
$c_f$	&	Pore fluid compressibility					\\
$\alpha_f$	&	Pore fluid bulk thermal expansion coefficient					\\
				
\end{tabular}

\section{Introduction}
\label{sec:ext:intro}

\begin{figure}
  \includegraphics[width=0.5\textwidth]{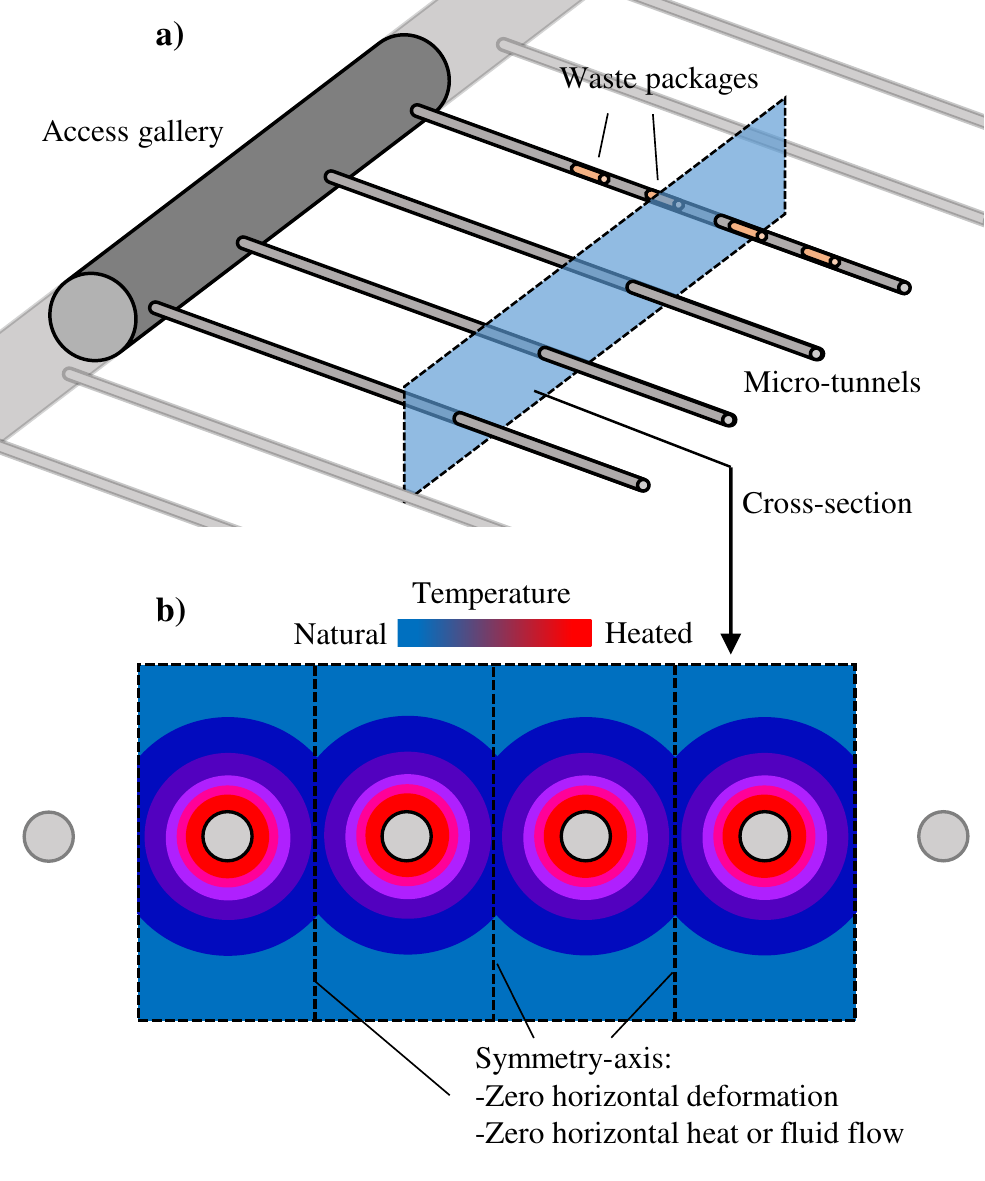}
\caption{a) Layout of the micro-tunnels containing high-level radioactive waste (adopted from \citealp{Andra2005}), with b) a cross-section where \hl{one} can assume plane strain and zero heat and temperature flux perpendicular to the image plane, illustrating the heat radiation from periodically constructed micro-tunnels.}
\label{fig:ext:context}      
\end{figure}

A detailed investigation of the thermo-poroelastic properties of host rocks is essential in the design of the deep geological disposals of high-level radioactive wastes in clay rocks, as considered in France, Switzerland and Belgium. \hl{Next} to the hydro-mechanical response involved during the excavation of the micro-tunnels, \hl{a thermo-hydro-mechanical response takes place during to the heat release from the exothermic waste packages} (e.g. \citealp{Gens200720,Seyedi201775,Armand201741,Conil2020}, see Fig. \ref{fig:ext:context}). Heat radiates from the packages, resulting in gradually increasing temperature in the rock. \hl{This so-called} thermal pressurization \hl{occurs} when saturated soils or rocks are heated in undrained conditions, due to the significant difference between the thermal expansion coefficients of water and of the solid phase \citep{Ghabezloo200901,Vu2020}. Quasi-undrained conditions are provided through the very low permeability of the claystone (in the order of 10$^{-20}$ m$^2$, \citealp{Escoffier200532,Davy200766,Menaceur201529,Menaceur201618}).

\hl{Thermal pressurization} has been observed in various in situ heating tests carried out in underground research laboratories (URL) excavated in claystones. It has been described and numerically modelled by different authors (e.g. by \citealp{Gens200720,Jobmann2007} in the HE-D test run in the Opalinus clay in the Mt Terri URL in Switzerland, by \citealp{Seyedi201775,Conil2020} in the TED test \hl{and by \mbox{\cite{Armand201741,Bumbieler2021,Tourchi2021}} in the ALC1604 test, both} run in the Callovo-Oxfordian claystone (COx) in the Bure URL in France). Thermal pressurization has also been investigated through laboratory experiments, as done by \citealp{Mohajerani201211,Zhang201746,Braun2020th} on COx claystone and \citealp{Monfared201173} on Opalinus clay. 

The thermo-poromechanical strain response of \hl{claystones} in drained and undrained conditions has been investigated in the laboratory by various authors including \citealp{Zhang201746,Mohajerani201413,Menaceur201618,Belmokhtar201787,Belmokhtar201722,Braun2020lett,Braun2020th} on the COx claystone and \citealp{Monfared201173} and \citealp{Favero2016} on the Opalinus clay. A thermo-elastoplastic behaviour upon drained heating under constant effective stress has been evidenced by \citealp{Monfared201173} on the Opalinus clay and by \citealp{Belmokhtar201722,Braun2020lett,Braun2020th} on the COx claystone. The latter authors found that \hl{both elastic and plastic} drained strains were anisotropic during \hl{the tests}. These features are somewhat similar to what has been observed in previous investigations on clays (e.g. \citealp{Baldi198880,Sultan200213} on Boom clay and \citealp{Abuel200714} on Bangkok clay).

Figure \ref{fig:ext:context}a illustrates schematically the current French concept for high \hl{level} radioactive waste disposal at great depth, in which waste canisters are deposited in parallel horizontal sleeved micro-tunnels of 0.7 \hl{to 1.0 m} diameter and 80 \hl{to 150} m length \hl{\mbox{\citep{Armand2015,Plua2021a,Plua2021}}}. The micro-tunnels are evenly spaced with enough distance such as to keep the temperature below 90 °C in \hl{the rock and to ensure no damage in the far field}. Given the length of the micro-tunnels and their periodic layout, thermo-hydro-mechanical processes are often modelled assuming plane strain conditions (Fig. \ref{fig:ext:context}b), with symmetry conditions at mid-distance between them. This results in peculiar conditions at this location, i.e. zero horizontal deformations and no horizontal heat nor fluid flux. One can expect \hl{a significant} thermal pore pressure build-up in this area, since the horizontal fluid flow is limited through the no-flux boundary conditions. \hl{With time,} thermal pore pressures dissipate towards the far-field. \hl{In summary,} the rock mass at this mid-distance is submitted to zero lateral deformation, constant overburden vertical total stress and temperature increase. Constrained lateral thermal expansion leads to an increase of lateral total stress. 
In such conditions, one has to consider the risk of reaching a pore pressure level larger than the total vertical stress, \hl{which} could lead to local vertical tensile failure \hl{\mbox{\citep{Li2018}}}. \hl{Note that the planned design of the high level radioactive waste repository does not allow any tensile failure. More importantly, the study of thermally induced fracturing is crucial for risk management.}

The tensile strength of clay-rocks has been widely investigated in rock mechanics, and commonly measured through the direct tension test \citep{Hoek196450,Brace196411} or the Brazilian test \citep{Akazawa1943,Carneiro1943,ISRM197831}. 
\cite{Yang199783} determined the tensile strength of a transversely isotropic claystone in direct tension tests. They evidenced a smaller tensile resistance for tension applied perpendicular to the plane of isotropy ($\sim 4$ MPa) with respect to that measured on specimens loaded parallel to the plane of isotropy ($\sim 11.5$ MPa). Interestingly, under tension loading, they found elastic material properties close to the values measured under compression. Only the Young modulus perpendicular to the plane of isotropy was detected to be higher in tension than in compression. They observed a quite linear stress-strain behaviour under tension. 
\cite{Coviello200525} carried out various different types of tensile tests, to determine the strength of two types of weak rocks, Gasbeton and calcarenite. In the case of calcarenite, they found the same tensile strength with both direct tensile test and Brazilian test ($\sim 0.6$ MPa). In the case of Gasbeton, the strength measured with direct tension ($\sim 0.9$ MPa) was higher than measured in Brazilian tests ($\sim 0.5$ MPa). They noted that for both materials, the Young moduli in tension and compression were about the same, with a very linear stress-strain behaviour until failure.
In the work of \cite{Hansen1987}, an overview of various shales \hl{and claystones} is given with their unconfined compressive strength, elastic properties and tensile strength, presented in Tab. \ref{tab:ext:criterion_literature}. These are compared with the data from \cite{Bossart201126} on the Opalinus clay.

According to \cite{Diederichs200710}, one can estimate $\sigma_t=\sigma_c/m_i$, \hl{where $\sigma_t$ is the tensile strength, $\sigma_c$ the unconfined compressive strength and $m_i$ a material constant of the Hoek-Brown model. An estimation for claystone of $m_i \approx 4.0 \pm 2$ is recommended by \mbox{\cite{Hoek200719}}.}
Even though \cite{Perras201452} concluded that \hl{this formula} does not give accurate results, \hl{it} can be used to get a first \hl{approximation} of the tensile resistance of brittle rocks. The calculated values \hl{for} $m_i$ of some shales \hl{and claystones} are presented in Tab. \ref{tab:ext:criterion_literature}, confirming that \hl{they} are generally higher than the recommended value of $4.0 \pm 2$. \hl{One observes} a rather large variability \hl{of $m_i$ for} these clay rocks between 6.5 and 13.6. \hl{Among the aforementioned studies, authors did not investigate a temperature dependency of the tensile resistance.} 

\begin{table*}
\caption{Shale \hl{and claystone} characteristics adopted from the literature, with an estimation of the Hoek-Brown coefficient $m_i=\sigma_c/\sigma_t$, compared with the experimental results of this study on COx claystone}
\begin{tabular}{llllll}
\hline\noalign{\smallskip} 
Rock type              & $\sigma_c$  & $E$    & $\nu$    & $\sigma_t$ & $m_i$      \\
			             &{[}MPa{]} & {[}GPa{]}    & {[}-{]}   & {[}MPa{]}  & {[}-{]}     \\
\noalign{\smallskip}\hline\noalign{\smallskip} 
Pierre shale$^1$      & 7.2  & 0.6  & 0.12 & 0.5  & 13.6    \\
Rhinestreet shale$^1$ & 58.7 & 17.5 & 0.19 & 8.4  & 7.0     \\
Green river shale$^1$ & 94.8 & 10.4 & 0.31 & 11.9 & 8.0    \\
Carlile shale$^1$          & 22.8 & 3.3  & 0.16 & 3.5  & 6.5    \\
Colorado shale$^2$    & 2.97 & 2.4  & 0.42 & 0.4  & 6.9   \\
Opalinus clay $\bot$ $^3$     & 25.6 & 2.8  & 0.33 & 2.0  & 12.8      \\
Opalinus clay $\|$ $^3$              & 10.5 & 7.2  & 0.24 & 1.0  & 10.5      \\
COx claystone $\|$ &17.8$^4$&5.7$^5$ &0.29$^5$&3.0$^6$	&5.15$^6$\\
\noalign{\smallskip}\hline\noalign{\smallskip} 
\multicolumn{6}{l}{$^1$\cite{Hansen1987},}\\
\multicolumn{6}{l}{$^2$\cite{Mohamadi201634}, $^3$\cite{Bossart201126}, }\\
\multicolumn{6}{l}{$^4$Andra database, $^5$\cite{Braun2020el}, $^6$this study }          
\end{tabular}
\label{tab:ext:criterion_literature}
\end{table*}

The possibility of fracturing \hl{claystones} through temperature\hl{-}induced pore pressures has been recently demonstrated by \cite{Li2018} through laboratory unconfined heating tests. In lack of any stress and strain measurements, these authors supposed that the fractures, that propagated parallel to the bedding plane, were induced when the pore pressure exceeded the tensile strength.
To the authors best knowledge, the thermal fracturing of geomaterials has hitherto not been investigated under \hl{laboratory} conditions of controlled stresses, pore pressure and temperature. This paper presents the development of a novel thermal extension triaxial device aimed at mimicking the thermal extension failure phenomenon \hl{under particular stress paths}. 
\hl{The test results show the THM response of the COx claystone from the intact to the fractured state when subjecting it to thermal pressurization}
and provide an appropriate criterion accounting for its thermally induced tensile failure.

\section{Novel experimental device}
\label{sec:ext:device}

This work's aim was to develop a thermal extension triaxial device, in which a specimen should be submitted in undrained conditions to a temperature increase. \hl{At the same time,} radial strain \hl{has to remain zero} and vertical total stress \hl{constant, which corresponds} to the in situ overburden stress. The zero radial stress condition involves a servo-control of the confining pressure, \hl{while} maintaining pore pressures larger than the total vertical stress is another experimental challenge.

\begin{figure*}
\centering
  \includegraphics[width=\textwidth]{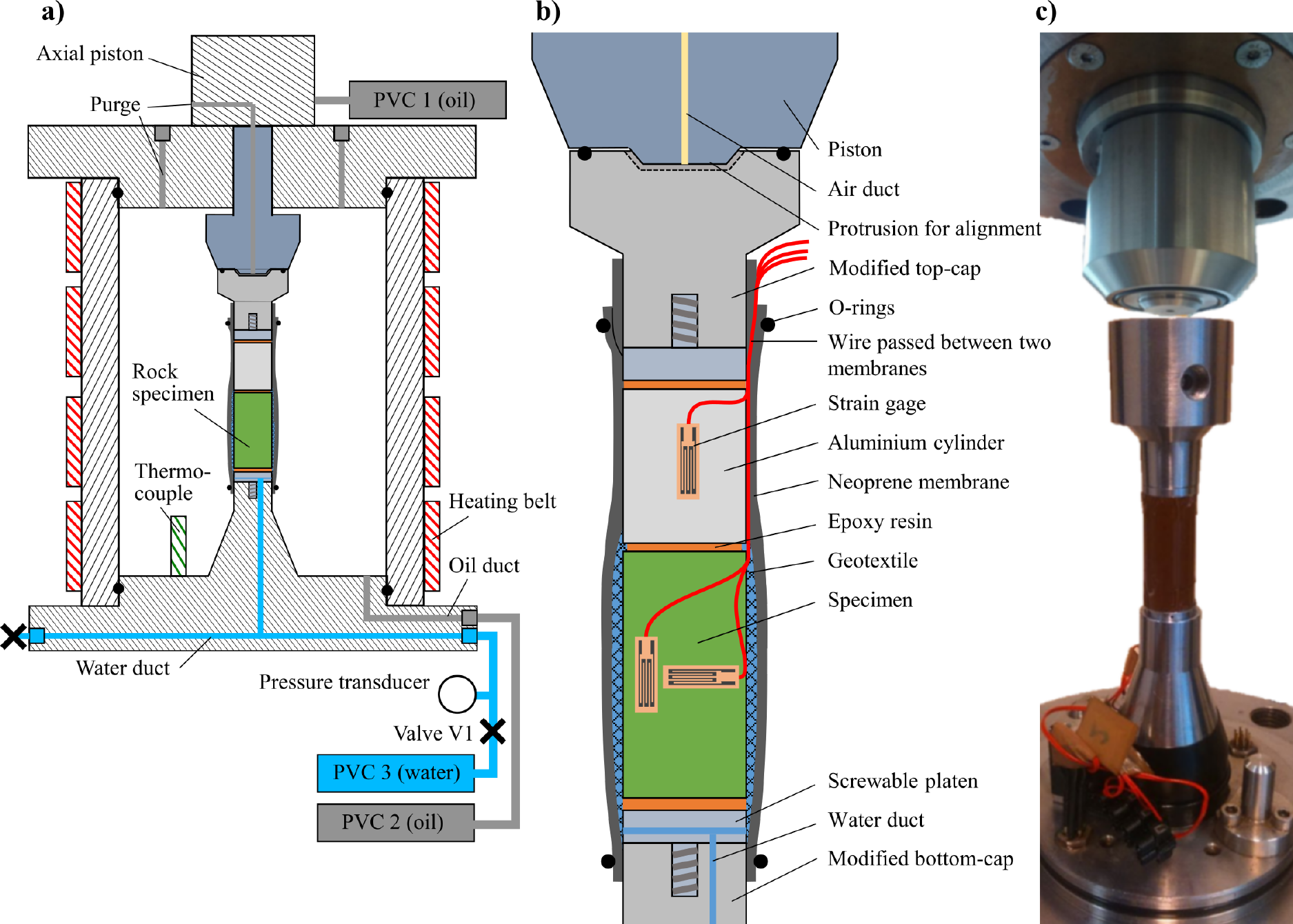}
\caption{a) Schematic view of the triaxial cell. b) Zoom on the specimen assembly with novel extension modifications, permitting to maintain pore pressures larger than the axial total stress through glued connections, while saturation and drainage is achieved through lateral geotextiles. c) Photo showing a dummy specimen mounted with bottom and top caps and the axial piston. The aluminium force transducer is not shown in the photo.}
\label{fig:ext:triax_ext}      
\end{figure*}

Similar to previous thermal testing devices developed and used in the same research group at Ecole des Ponts ParisTech (e.g. \citealp{Menaceur201618,Belmokhtar201722,Braun2020th}, a triaxial cell (Fig. \ref{fig:ext:triax_ext}a) was employed for the laboratory tests. A pressure volume controller (PVC1, GDS brand) was used to apply pressure to the axial piston and \hl{to} control the axial stress. Additional PVCs (PVC2,3, GDS brand) were used to control the confining pressure and the pore pressure separately. Temperature changes were applied by using an electric silicone heating belt, wrapped around the steel cell, and measured by means of a thermocouple located inside the cell. Local (radial and axial) strains were accurately monitored by means of strain gages glued directly onto the sample (see \citealp{Braun2019}). Accuracy in radial strain measurement is here of utmost importance so as to ensure satisfactory zero lateral strain during heating.

\subsection{Zero radial strain condition}
\label{sec:ext:strain_control}

The zero radial strain condition was achieved \hl{through} a servo-controlling routine programmed in the LabView software. \hl{The routine couples the strain measurements and the cell pressure PVC.} Since the cell temperature changes during a test, it is required to correct the measured strains in real-time for thermally induced measurement errors. Following a correction method presented by \cite{Braun2019}, a reference gage on a piece of steel 316L \hl{was} placed in the triaxial cell next to the tested specimen. Doing so, thermal and mechanical strains on a reference material \hl{can be recorded}. 
By knowing the thermal expansion coefficient and compressibility of steel 316L, the errors induced by any changes in cell temperature, room temperature, and confining pressure, can be determined as the difference between the measured reference strain and the calculated steel strain. This error was then subtracted from the strain measurements on the specimen. \cite{Braun2019} checked the quality of the thermal strain measurements by testing a specimen of aluminium alloy 2011 under thermal loads. A linear thermal expansion coefficient of the aluminium specimen of $2.30 \times 10^{-5}$ $^{\circ} \mathrm{C}^{-1}$ was evaluated after correction, which fairly well corresponded with that provided by the manufacturer. 

During the laterally constrained heating tests, the corrected lateral strains \hl{were monitored}, which were maintained at zero value through a partial-integral-derivative (PID) software controller. This servo-control routine calculated the required lateral stress increment, which was then transmitted to the PVC to change the confining stress accordingly. The pressure applied by the axial PVC was simultaneously controlled in order to maintain constant the axial stress. Tuning of the PID controller parameters was carried out before the tests. It was possible to adapt them during the tests, so as to ensure a fast response of the confining PVC and to avoid overshoot and oscillations. The PID routine was executed in intervals of 30 seconds, which proved being sufficiently fast with respect to the adopted heating rate of maximum 10 $^\circ$C/h. The derivative part of the PID was not necessary to achieve precise control.

\subsection{Application of tensile axial stress}
\label{sec:ext:axial_stress}

The main issue met in tension tests in triaxial cells for soils and rocks \hl{is the loss of} contact between the axial piston and the top cap, \hl{once} the axial stress becomes smaller than the radial one (i.e. the confining pressure). \cite{Donaghe1988} presented a review of the different available systems to avoid this loss of contact in the case of soil testing. They mention screw connections, resin pot systems or a suction cap. The former two methods are difficult to implement when the piston is not easily accessible once the contact with the top cap is made. The suction cap, however, allows closing the cell and moving the piston downwards. Once in contact, a certain surface between the top cap and piston is isolated from the confining stress by means of a rubber membrane. By ensuring a lower pressure than the cell pressure within this surface, and given that the isolated surface is larger than the specimen cross section, the confining pressure keeps the two pieces compressed together. \hl{This allows one} to transmit tension \hl{to the specimen}.  
In \hl{the presented} system, the connection between the piston and the top cap was based on the principle of a suction cap, by using a flat disc shaped volume isolated by means of an O-ring and connected to atmospheric pressure (Fig. \ref{fig:ext:triax_ext}b,c). \hl{As required,} the isolated volume \hl{has a} base surface larger than the specimen cross section. 
As seen in Fig. \ref{fig:ext:triax_ext}b,c, a protrusion in the piston ensures alignment of cap and piston.

Another issue is the loss of contact between top cap and specimen, once the pore pressure becomes equal to the total axial stress. Pore pressures higher than the axial stress cannot be maintained in a device without adequate modifications. 
A comprehensive overview of different methods to overcome this problem has been given by \cite{Perras201452}. They recalled that the two main techniques for direct tensile testing consisted either in attaching the specimen ends to the load frame (by using grips or by gluing, cf. \citealp{Fairhurst196110,Hawkes197017}), or in modifying the sample shape to a so-called dog-bone shape \citep{Hoek196450,Brace196411}. The latter method is not recommended for weak rocks such as claystones, as these materials would not withstand the required shaping process in lathe \citep{Perras201452}. The former method of gluing or gripping specimen has the advantage that conventional cylindrical specimens can be used. \cite{Hawkes197017} have shown that strain inhomogeneities could be reduced, when \hl{the specimen was glued only on its end surfaces and not} along lateral sections. It was still likely that failure occurred close to the glued end surfaces.
\hl{In the presented device, the specimen and the caps were therefore glued together} by tensile-resistant epoxy resin. 

Moreover, an internal axial force transducer \hl{was integrated} between the specimen and the top cap. Given that no convenient 20 mm diameter force gage was commercially available, a cylindrical aluminium (alloy 2011) piece of 30 mm height and 20 mm diameter was equipped with an axial strain gage (Fig. \ref{fig:ext:triax_ext}b). This force transducer was previously calibrated, taking into account its thermal dilation. 

In the final assembly (Fig. \ref{fig:ext:triax_ext}b), the aluminium cylinder was glued on top of the specimen. Afterwards, two specially designed screw-able disk-shaped steel platens were glued onto the top of the force transducer and onto the bottom end-surface of the specimen. A two-component epoxy resin (Araldite 2014, Huntsman brand) \hl{was used}, with a tensile strength of 26 MPa, much larger than the expected strength of the tested specimens. Its glass transition temperature, up to which the resin maintains its original strength, is 80°C. \hl{This} temperature that was not exceeded during the tests. Once the resin hardened, the assembly was screwed to the bottom cap, which was fixed to the triaxial cell. The top cap was also screwed onto the assembly, allowing us to maintain tensile stresses. The drainage lines drilled into the bottom platten were connected via the top and bottom cap to a PVC that controls the pore pressure. Since the top and bottom surfaces of the specimen have been made impermeable by impregnating with epoxy resin, drainage had to be ensured by means of a geotextile wrapped around the specimen. The geotextile was connected to the lateral drainage line of the top and bottom platens. \cite{Monfared201173} have verified that these geotextiles remain permeable even under high confining pressure. A neoprene membrane was then put over the specimen and the geotextile, isolating the confining fluid (silicone oil) from the pore fluid.

The system was developed for specimens with a reduced diameter of 20 mm so as to enable faster drainage and saturation (see \citealp{Belmokhtar201819}). \hl{The drainage length corresponds} to the specimen radius of 10 mm \hl{and} the height of the specimens was close to 35 mm.

\subsection{Calibration of the drainage system}
\label{sec:ext:calib}

\cite{Ghabezloo201060} summarized the different properties of the drainage system, which are the volume of the drainage system $V_L$, its compressibility $c_L$ due to pressure changes in the drainage system $p_L$, the compressibility $\kappa_L$ due to radial confining pressure changes $\mathrm{d}\sigma_\mathrm{rad}$, and the bulk thermal expansion coefficient $\alpha_L$ under a temperature change $\mathrm{d}T$. The change in volume of the drainage system can hence be expressed as:
\begin{equation}
\label{eq:ext:dV_L/V_L}
\frac{\mathrm{d}V_L}{V_L}=c_L \mathrm{d}p_L+\alpha_L \mathrm{d}T-\kappa_L \mathrm{d}{\ \sigma}_\mathrm{rad}
\end{equation}
with the different parameters defined as:
\begin{equation}
\label{eq:ext:c_L}
c_L=\frac{1}{V_L}\left(\frac{\partial\ V_L}{\partial\ p_L}\right)_{T,{\ \sigma}_\mathrm{rad}}
\end{equation}
\begin{equation}
\label{eq:ext:alpha_L}
\alpha_L=\frac{1}{V_L}\left(\frac{\partial\ V_L}{\partial\ T}\right)_{\ p_L,{\ \sigma}_\mathrm{rad}}
\end{equation}
\begin{equation}
\label{eq:ext:kappa_L}
\kappa_L=-\frac{1}{V_L}\left(\frac{\partial\ V_L}{\partial\ {\ \sigma}_\mathrm{rad}}\right)_{T,\ p_L}
\end{equation}

Pseudo-undrained conditions are here defined as a constant sum of fluid mass $\mathrm{d} M_f=\mathrm{d} M_s+\mathrm{d} M_L=0$ (where $ M_s = m_f V_s$ is the fluid mass in the specimen, $m_f$ the specimen fluid mass per unit volume, $V_s$ the specimen total initial volume, $M_L = V_L \rho_L$ the fluid mass in the drainage system and $\rho_L$ the density of the fluid in the drainage system). Rewriting the equations of \cite{Ghabezloo201060}, \hl{one obtains}:
\begin{equation}
\label{eq:ext:drainage_balance}
V_L\rho_f\left[\left(c_L+c_f\right)\mathrm{d}p_L+\left(\alpha_L-\alpha_f\right)\mathrm{d}T-\kappa_L \mathrm{d}{\sigma}_{rad}\right]
+V_s\mathrm{d}m_f=0
\end{equation}
where $\rho_f$, $c_f$ and $\alpha_f$ are the \hl{pore}fluid density, compressibility and bulk thermal expansion coefficient, respectively.
In \hl{the presented} experimental configuration \hl{drainage occurs} through the lateral specimen surface, therefore the fluid pressure in the drainage system $p_L$ is equal to the specimen pore pressure on this surface. 

Calibration tests \hl{were carried out} to characterize the properties of the drainage system, assuming that they are constant with stress, pore pressure and temperature. To this end, a steel dummy specimen with zero porosity ($\mathrm{d}M_s=0$) \hl{was installed}. Saturating the empty drainage system with water provided $V_L=3500$ {mm}$^3$. Pore pressure changes \hl{were applied}, while measuring $\mathrm{d}V_L$ with PVC3, which allowed us to calculate $c_L=1.33$ GPa$^{-1}$. 
Changing the radial stress and measuring the pore pressure change in a closed drainage system \hl{resulted in} $\kappa_L=0.44$ GPa$^{-1}$. The parameter $\alpha_L=0.79\times$10$^{-4}$ °C$^{-1}$ was determined by recording pore pressure changes during temperature cycles.


\section{Specimen characterization and preparation}
\label{sec:ext:material}

The specimens of the COx claystone come from horizontal cores (EST 53650 and EST 57185) extracted at a depth of 490 m in the Bure URL. The clay content of the COx claystone at this level is around 42 \%, with an average porosity of 17.5 \% and an average water content of 7.9 \% \citep{Robinet2012,Conil201861}. The in situ stress conditions were measured by \cite{Wileveau200786}, with a vertical and a minor horizontal total stress of around 12 MPa, a major horizontal stress of around 16 MPa and a hydrostatic pore pressure of about 4.9 MPa.

Avoiding the desaturation and mechanical damage of cores is an important concern, since the claystone mechanical properties \hl{should be kept} as close as possible to the natural ones. Shales and claystones are known to be particularly sensitive to changes in water content \citep{Chiarelli2003,Vales2004,Zhang2004,Pham200764,Zhang201279}, that may occur during the successive processes of coring, storage, transportation and trimming of laboratory specimens \citep{Chiu1983,Monfared201163,Ewy201538,Wild2017}. Great attention has hence to be paid to preserve the initial water content of the cored specimens. To avoid any contact with water, Andra is used to carry out coring with air-cooling, prior to sealing the extracted cores in so-called T1 cells \citep{Conil201861}. T1 cells accommodate a 320 mm long COx core with 80 mm diameter, previously wrapped in aluminium foil and in a latex membrane to avoid drying. A larger diameter PVC tube is placed around the core and cement is cast in the annulus between the membrane and the tube, creating a rigid mechanical protection. After cement hardening, a metal spring is used to constrain the core along the axial direction.

Once arrived in the laboratory, T1 cells were opened and the cores were immediately covered by a layer of paraffin wax, so as to prevent drying during the trimming process. Cylindrical proofs of 20 mm diameter were drilled perpendicular to the bedding plane by means of a diamond coring bit cooled by compressed air. 
\hl{Note that according to Ba{\v{z}}ant's theory on the size effect in fracture mechanics, the strength of a rock specimen decreases with its structural size. Moreover, there is the possibility of a statistical size effect, meaning that the probability for encountering heterogeneities of low strength increases with increasing sample size \mbox{\citep{Bazant1984}}. Such effects couldn't be identified in this study since only 20 mm diameter samples were tested.} 
\hl{All samples were oriented with their axis perpendicular to bedding. This facilitates the application of the in situ condition of zero horizontal strain (along the bedding plane) and constant overburden pressure perpendicular to bedding. Such in situ condition favours the tensile failure along the weak bedding plane. This orientation has also the particularities that: a) The tensile fracture direction imposed by the experimental conditions coincides with the bedding plane. b) Radial strains in the direction parallel to bedding are uniform. 
If the sample is cored along the bedding plane, it would deform non-uniformly in radial directions in triaxial tests, due to thermo-poro-elastic anisotropy. In this case, one would need three orientations of strain gages: axial parallel to the bedding plane, radial parallel to the bedding plane, and radial perpendicular to the bedding plane. Zero radial strain conditions in this configuration would only be possible, if the radial anisotropy of thermal dilation and of mechanical contraction is the same.} 
\hl{In the next step,} the cylinders were cut at desired length (30 - 40 mm) by using a diamond string saw, to obtain parallel end surfaces. The specimens were afterwards wrapped in an aluminium foil and covered by a mixture of 70 \% paraffin and 30 \% vaseline oil, and stored until running the tests. Small cuttings of the cores were taken after opening the core, to conduct a petrophysical characterisation (Tab. \ref{tab:ext:1}). The volume of the cuttings was measured by hydraulic weighting in petroleum, while the dry density was obtained after oven drying 24 h at 105 $^\circ$C. A solid density of 2.69 Mg/m$^3$, provided by Andra, was adopted to calculate the porosity $\phi$ and the degree of saturation $S_r$. The specimen suction $s$ was determined by using a chilled mirror tensiometer (WP4C, Decagon brand). As seen in Tab. \ref{tab:ext:1}, the high degrees of saturation (92.5 \% and 95.3 \%) and low corresponding suction values (24.2 and 17.4 MPa) indicate a good conservation of the cores and a satisfactory sample quality.

\begin{table*}
\caption{Mean and standard deviation (in brackets) of petrophysical measurements done on cuttings of two COx cores.}
\label{tab:ext:1}       
\begin{tabular}{ lccccccc }
	\hline\noalign{\smallskip}
  	\multirow{2}{*}{Core}	&\multirow{2}{*}{\#}& {$\rho$} & {$\rho_d$ } & {$\phi$} & {$w$} & {$S_r$}	& {$s$}\\
  							&& $[\text{g/cm}^3]$ & $[\text{g/cm}^3]$  & $[\%]$ & $[\%]$ & $[\%]$	& $[\text{MPa}]$\\
	\noalign{\smallskip}\hline\noalign{\smallskip}
 	{EST} 		&\multirow{2}{*}{1}	& 2.37 		& 2.22 		& 17.9 		& 7.5 		& 92.5 		& 24.2\\
 	{53650}		&& (0.00) 	& (0.01) 	& (0.2)  	& (0.1) 	& (0.8)		& (2.1)\\
 	\noalign{\smallskip}\hline\noalign{\smallskip} 
 	{EST} 	&\multirow{2}{*}{2}	& 2.38 		& 2.21 		& 18.2 		& 7.9 		& 95.3 		& 17.4 \\
 	{57185}	&	& (0.00) 	& (0.00) 	& (0.2)  	& (0.1) 	& (0.7)		& (0.1)	\\
\noalign{\smallskip}\hline
\end{tabular}
\end{table*}

\begin{figure}[htbp]
\includegraphics[width=0.5\textwidth]{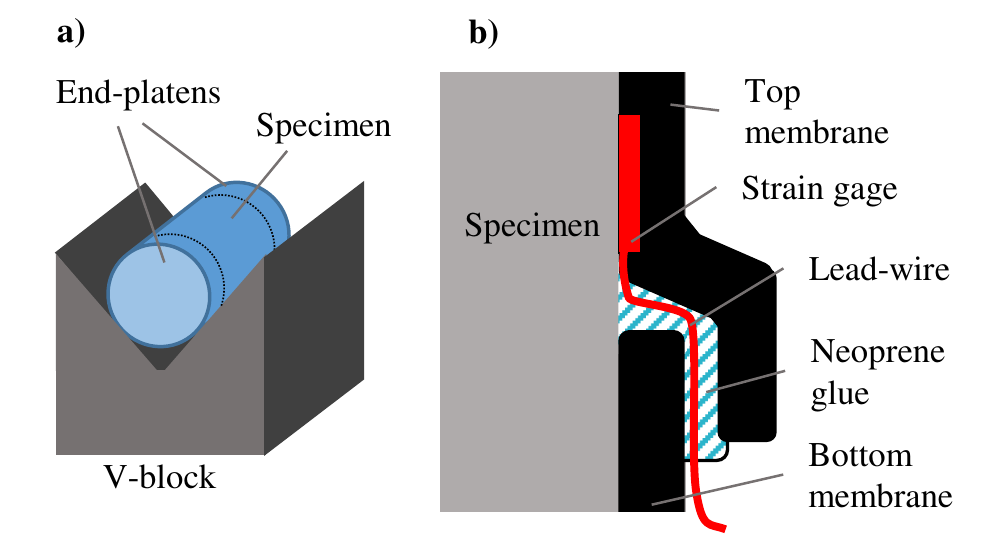}
\caption{a) Attaching the end platens to the specimen using epoxy adhesive, placed in a V-block to ensure axial alignment; b) schematic cut to illustrate the layout of two overlapping membranes to pass-through strain gage wires.}
\label{fig:ext:preparation}       
\end{figure}

Once removed from storage and unwrapped from aluminium foil and paraffin, the lateral surfaces of the specimen were protected from drying by wrapping them in an adhesive tape. To ensure correct axial alignment between the specimen and the end platens, the specimen was placed in a V-block with clamp holders (Fig. \ref{fig:ext:preparation}a) and a layer of epoxy resin was applied to both end surfaces. It was observed that the epoxy resin bonded very well to the claystone, provided that the sample surfaces were previously cleaned from dust by wiping with a paper towel. Bonding the epoxy resin to the surfaces of the metal platens required more caution: the surfaces had to be previously roughened with coarse sand paper and wiped with acetone to remove any dirt or grease. The top and bottom platens were affixed \hl{to the specimen} with the resin and fastened with clamp holders. To achieve full bonding at room temperature, the epoxy resin had to cure for about {20 h}. The assembly was then taken from the V-block, the adhesive tape removed and the two strain gages glued to the specimen at mid-height, along the axial and radial directions (Fig. \ref{fig:ext:triax_ext}b). A layer of geotextile was wrapped around the specimen, while ensuring contact between the geotextile and the outlets integrated in the bottom platen. Two neoprene membranes were superposed over the geotextile, as indicated in Fig. \ref{fig:ext:preparation}b. The strain gages wires were passed through a section where the membranes overlapped, and sealed with neoprene glue.

In order to prevent any contact of the sample with water before applying the confining stress, to avoid excessive swelling, the drainage ducts were flushed with air. The specimen was installed in the triaxial cell and connected to the top and bottom caps, by using the screw connections on the two end platens. O-rings were put around the membrane to prevent any leaks on both top and bottom caps. Once the strain gage wires were connected to the data acquisition system, the cell was closed, filled with silicone oil and wrapped by the heating belt to enable temperature control.

\section{Experimental results}
\label{sec:ext:results}

\hl{In the following sections, the notation of Therzaghi effective stress is used to analyse stress states during experiments. The effective stress $\sigma'_i$ in a direction $i$ is obtained by $\sigma'_i = \sigma_i - p_f$, where $\sigma_i$ is the total stress and $p_f$ the pore pressure.}
A seen in Tab. \ref{tab:ext:results}, three tests were conducted under different levels of radial effective stress so as to investigate the effect of radial effective stress on the THM stress-strain behaviour and thermal failure of the COx claystone. Specimen EXT1 was tested starting from an isotropic effective stress state close to the in situ one ($\sigma_{z}'$ = 7.7 MPa, $\sigma_{h}'$ = 6.9 MPa and pore pressure $p_f$ = 4.9 MPa). EXT2 was started at a lower isotropic effective stress ($\sigma_{z}'$ = 2.7 MPa, $\sigma_{h}'$ = 3.5 MPa and $p_f$ = 2.1 MPa), while EXT3 was started at reduced axial effective stress but similar radial effective stress as EXT1 ($\sigma_{z}'$ = 3.2 MPa, $\sigma_{h}'$ = 8.4 MPa and $p_f$ = 2.1 MPa). 

\begin{table*}[]
\caption{Properties of the specimens measured during the experimental programme, where $\sigma$ is the isotropic confining stress, $\varepsilon_{\mathrm{hyd}}$ is the swelling strain during hydration, $\gamma_t$ the measured fracture angle with respect to the horizontal bedding plane.}
\label{tab:ext:results} 
\begin{tabular}{@{\extracolsep{3pt}}llcccccccccccc@{}}
\hline\noalign{\smallskip} 
       &      & \multicolumn{2}{l}{Saturation}                            & \multicolumn{4}{l}{Initial state} & \multicolumn{6}{l}{Failure} \\
\cline{3-4}\cline{5-8}\cline{9-14}\noalign{\smallskip}
& & $\sigma$ & $\varepsilon_{\mathrm{hyd}}$ & $T$             & $\sigma'_{z}$ & $\sigma'_{h}$ & $p_f$  & $T$             & $\sigma'_{z}$ & $\sigma'_{h}$ & $p_f$   & $\varepsilon_{z}$ & $\gamma_t$ \\
Sample & Core  & {[}MPa{]}              & {[}\%{]}                  & {[}$^{\circ}$C{]}            & {[}MPa{]}               & {[}MPa{]}               & {[}MPa{]} & {[}$^{\circ}$C{]}       & {[}MPa{]}  & {[}MPa{]}  & {[}MPa{]}   & {[}\%{]}                   & {[}$^{\circ}${]}              \\
\noalign{\smallskip}\hline\noalign{\smallskip} 
EXT1   & 2   & 8                & 0.2                 & 25.0          & 7.7              & 6.9               & 4.9 & 61.9    & -3.6 & 5.5  & 19   & -0.36                & 2              \\
EXT2   & 2   & 5                & 0.21               &  25.0          & 2.7              & 3.5               & 2.1 & 53.4    & -2.4 & 4.8  & 7.9  & -0.20                & 18             \\
EXT3   & 1   & 12               & 0.64                & 35.0          & 3.2              & 8.4               & 4.0 & 63.5    & -3.0 & 9.6  & 7.8  & -0.33                & 7               \\
\noalign{\smallskip}\hline           
\end{tabular}
\end{table*}

In tests EXT1 and EXT2, the cell was first brought to a constant initial temperature of 25°C, whereas tests EXT3 was brought at 35°C. The specimens were submitted to the target isotropic confining stress (5, 8 and 12 MPa, respectively, see Tab. \ref{tab:ext:results}) at a loading rate of 0.1 MPa/min. The drainage lines were previously dried and put under a vacuum of -80 kPa. Once the target confining pressure reached, the drainage lines were saturated with a synthetic pore water under a pressure of 100 kPa. The composition of this synthetic pore water was similar to that of the natural COx pore water, according to a composition provided by Andra,  
\hl{which consists of 1.95 g NaCl, 0.13 g NaHCO, 0.04 g KCl(2H$_2$O), 0.63 g CaSO$_4$(7H$_2$0), 1.02 g MgSO$_4$(2H$_2$O), 0.08 g CaCl$_2$ and 0.7 g Na$_2$SO$_4$ per litre of water.}
The applied back pressure of 100 kPa was chosen small enough to ensure a negligible decrease in effective stress and to minimize the poroelastic response of the claystone. Swelling strains due to hydration stabilized after about two days, with a maximum value \hl{$\varepsilon_{hyd} = 0.64 \%$} observed for sample EXT3 under 12 MPa (see Tab. \ref{tab:ext:results}). This order of magnitude is comparable to that observed by \cite{Belmokhtar201787,Braun2020el}. After hydration, the piston was brought in contact with the specimen and the desired initial values of the lateral stress ($\sigma_{h}'$), axial stress ($\sigma_{z}'$) and pore water pressure ($p_f$) were applied (see Tab. \ref{tab:ext:results}). \hl{It is} assumed that hydraulic equilibrium was achieved when reaching stable deformations.

\begin{figure}
\includegraphics[width=0.5\textwidth]{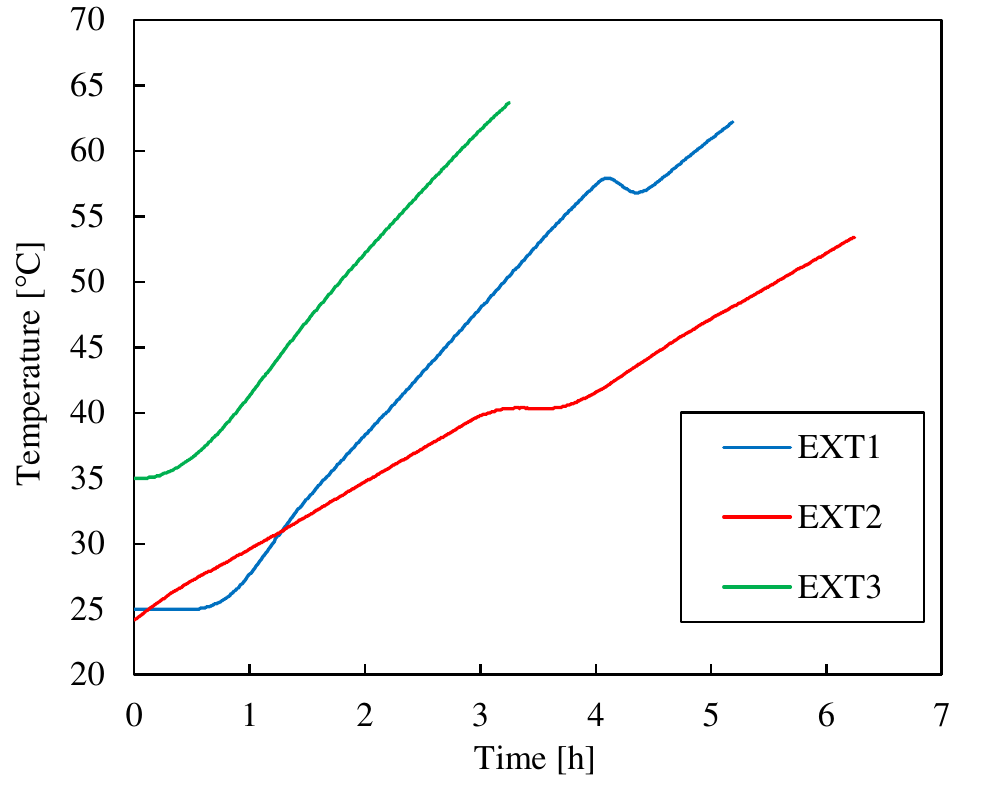}
\caption{Imposed temperature changes with respect to time, applied during the heating tests.}
\label{fig:ext:heatingpaths_ext}       
\end{figure}

The thermal extension tests were started by closing the drainage valve V1 (Fig. \ref{fig:ext:triax_ext}a) and heating the cell with a constant rate of to 10 $^\circ$C/h for samples EXT1 and EXT3, and 5 $^\circ$C/h for sample EXT2. Fig. \ref{fig:ext:heatingpaths_ext} shows the relatively constant rates of temperature elevation, while, due to some technical difficulties in the heating system, heating stopped for a brief time in experiments EXT1 after 4.1 hours and EXT2 after 3.2 hours. In both cases, the desired rate was recovered afterwards. During heating, the servo-control of the confining pressure was switched on to maintain radial strains at zero, while the hydraulic pressure in PVC1 was programmed to keep a constant vertical stress.

\begin{figure*}
\centering
  \includegraphics[width=\textwidth]{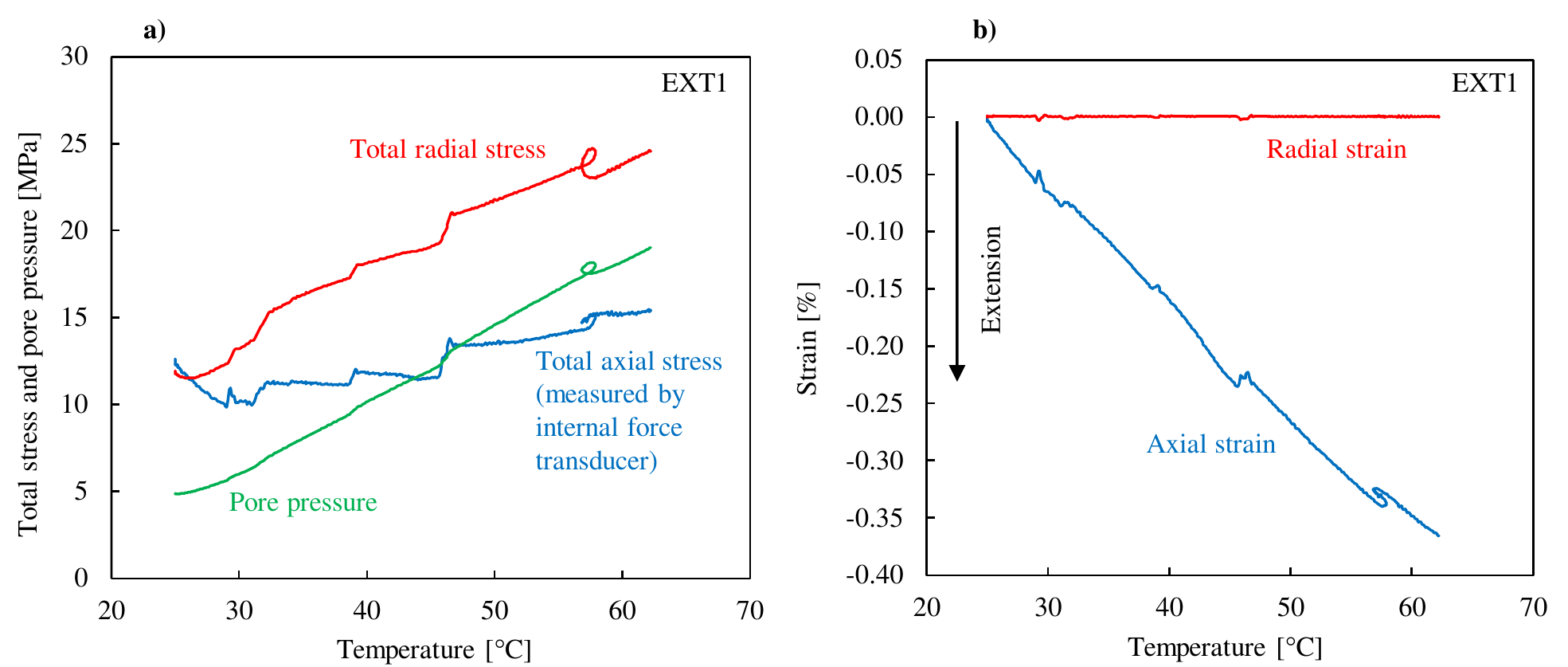}
\caption{Test EXT1: a) Total axial stress (measured by force sensor gage and desired to be kept constant by PVC1), total radial stress (servo-controlled by PVC to constrain zero radial strain) and thermally induced pore pressure. b) Radial strains which were maintained null by servo-controlling the confining pressure, and axial strains showing a specimen extension.}
\label{fig:ext1}      
\end{figure*}

Figure \ref{fig:ext1}a presents the data of test EXT1 in terms of applied total stress changes and measured pore pressure changes, with respect to increased temperature. The resulting measured strains in axial and radial direction are plotted in Fig. \ref{fig:ext1}b.
\hl{The data} confirms that the servo-control of zero radial strain could be achieved in the novel device. 
Note that due to the friction of the sealing gaskets within the axial piston housing, the axial stress on the specimen did not remain constant, even though the pressure applied by PVC1 (Fig. \ref{fig:ext:triax_ext}a) was stable. Nevertheless, these axial stress changes were recorded by the integrated aluminium force transducer \hl{(}except in EXT2, where the transducer failed\hl{)}. The direct measurement from the internal force transducer shows some irregularities and step changes between a minimum axial stress of 10.0 MPa and a maximum value of 15.5 MPa from the initial value of 12.6 MPa. \hl{This results in} an uncertainty of approximately $\pm 3$ MPa on the axial stress measured by PVC1. These stress variations are most likely due to the changes in radial stress and temperature, which affect the movement or static rest of the axial piston, given by the friction of its sealing components.
Unsurprisingly, it is observed that an increase in radial stress is necessary to keep the radial strain \hl{to zero during heating}. \hl{Increasing radial stress compensates the lateral} thermal expansion of the saturated specimen. Slight irregularities observed in the axial extension strain (Fig. \ref{fig:ext1}b) are a consequence of those observed in the axial strain control (Fig. \ref{fig:ext:triax_ext}a). \hl{Note that} the changes in axial strain are fairly linear \hl{until the maximum extension (–0.36 \%)} at the temperature at which thermal failure occurred (61.9 $^\circ$C). A linear increase in pore pressure from 4.9 MPa to a maximum of 19.0 MPa at 61.9 $^\circ$C is also observed (Fig. \ref{fig:ext1}a, Tab. \ref{tab:ext:results}).
Interestingly, all curves show a loop corresponding to the short temperature drop that occurred at 58 $^\circ$C.

\begin{figure*}
\centering
  \includegraphics[width=\textwidth]{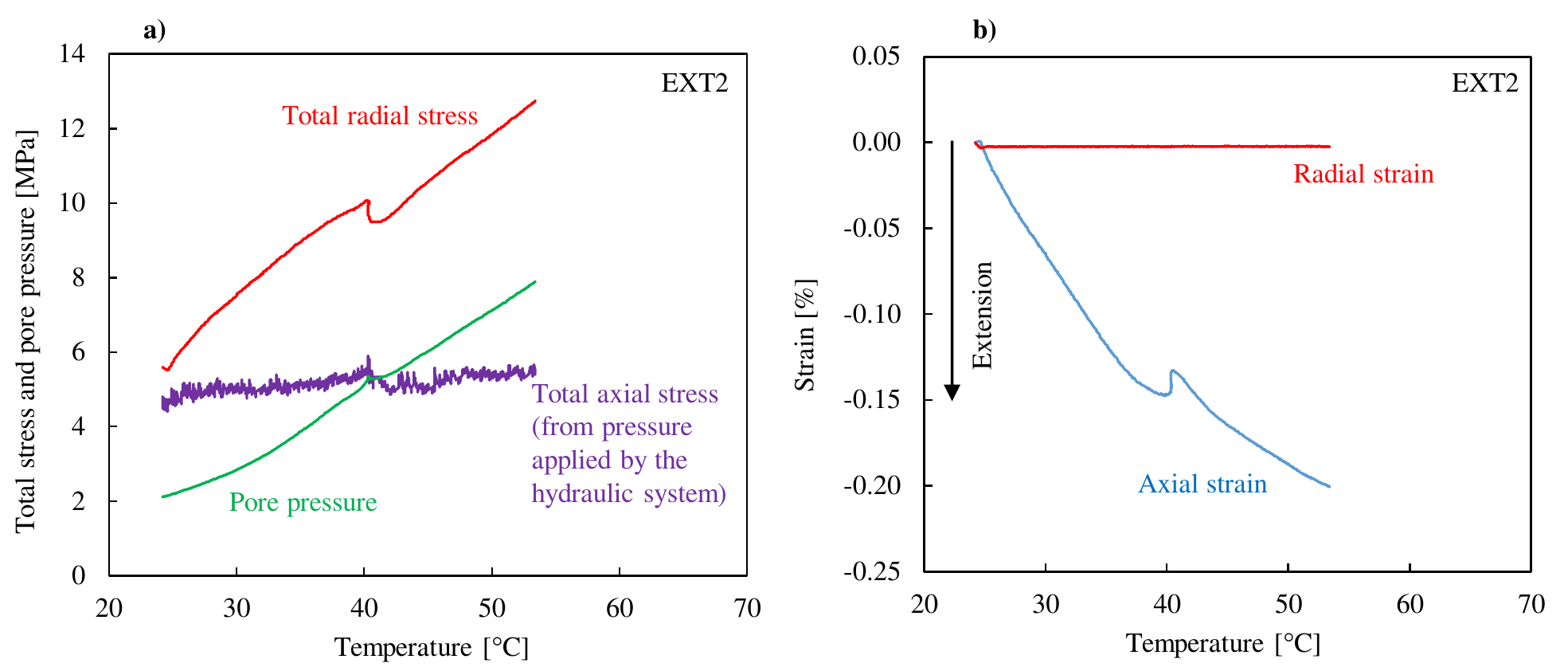}
\caption{Test EXT2: a) Total axial stress (measured by PVC1 due to failure of the axial force sensor gage), total radial stress (servo-controlled by PVC to constrain zero radial strain) and thermally induced pore pressure. b) Radial strains which were maintained null by servo-controlling the confining pressure, and axial strains showing a specimen extension.}
\label{fig:ext2}      
\end{figure*}

The control of zero radial strains was also ensured in test EXT2 (Fig. \ref{fig:ext2}). In this test, the internal axial force sensor malfunctioned, therefore only the less precise external axial stress measurement obtained by PVC1 \hl{are shown}. \hl{Note that this axial stress includes} the significant uncertainty of $\pm 3$ MPa. 
The maximum temperature at thermal failure was 53.4 $^\circ$C, with a maximum axial extension \hl{at $\varepsilon_z=$ –0.20 \%}. The pore pressure increased fairly linearly from 2.1 MPa to a maximum of 7.9 MPa (Fig. \ref{fig:ext2}a, Tab. \ref{tab:ext:results}).
The plateau observed at 40.5 $^\circ$C in the applied temperature (Fig. \ref{fig:ext:heatingpaths_ext}) caused a perturbation in the otherwise linear behaviour. The nonlinear temperature evolution produced a temporary decrease in total radial stress, axial compaction and pore pressure stabilization, followed by a recovery of the rates observed before the perturbation.

\begin{figure*}
\centering
  \includegraphics[width=\textwidth]{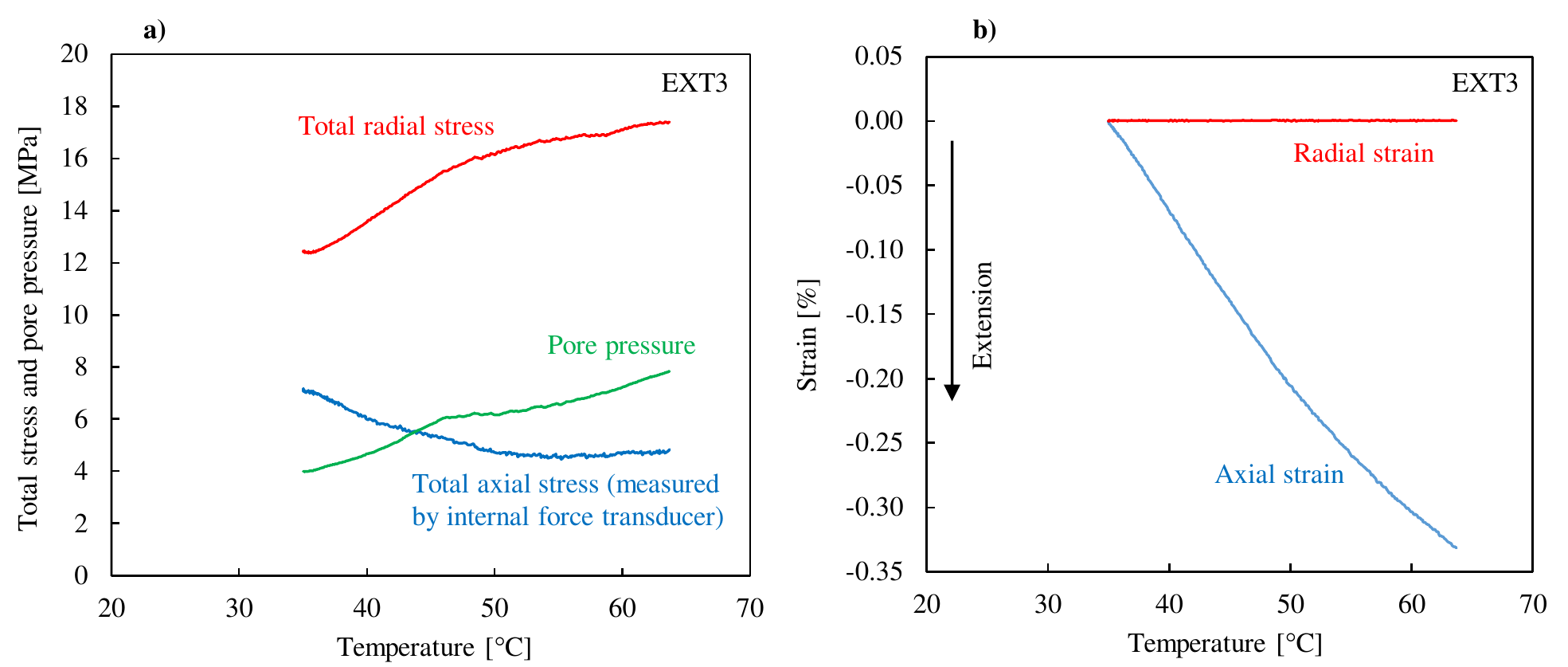}
\caption{Test EXT3: a) Total axial stress (measured by force sensor gage and desired to be kept constant by PVC1), total radial stress (servo-controlled by PVC to constrain zero radial strain) and thermally induced pore pressure. b) Radial strains which were maintained null by servo-controlling the confining pressure, and axial strains showing a specimen extension.}
\label{fig:ext3}      
\end{figure*}

In experiment EXT3 (Fig. \ref{fig:ext3}), one can observe a decrease in total axial stress from the initial value of 7.2 MPa down to 4.6 MPa at 53 $^\circ$C, followed by a stabilization with a slight increase above 53 $^\circ$C, up to 63.5 $^\circ$C, at which thermal pressurization failure occurred (Tab. \ref{tab:ext:results}). The axial stress variation lies within the uncertainty of the external control system due to piston friction ($\pm 3$ MPa). A clear coupling between the axial and the radial stress is again observed. The increase in pore pressure is \hl{almost} linear\hl{,} like in test EXT1.
A plateau in pore pressure changes is observed between 47 and 51 $^\circ$C, which is supposed to be due to non-linear material behaviour. Zero radial strain was again \hl{achieved}, while the axial strains decreased down to \hl{$\varepsilon_z=$} -0.34 \%. 

\begin{figure*}
\centering
  \includegraphics[width=\textwidth]{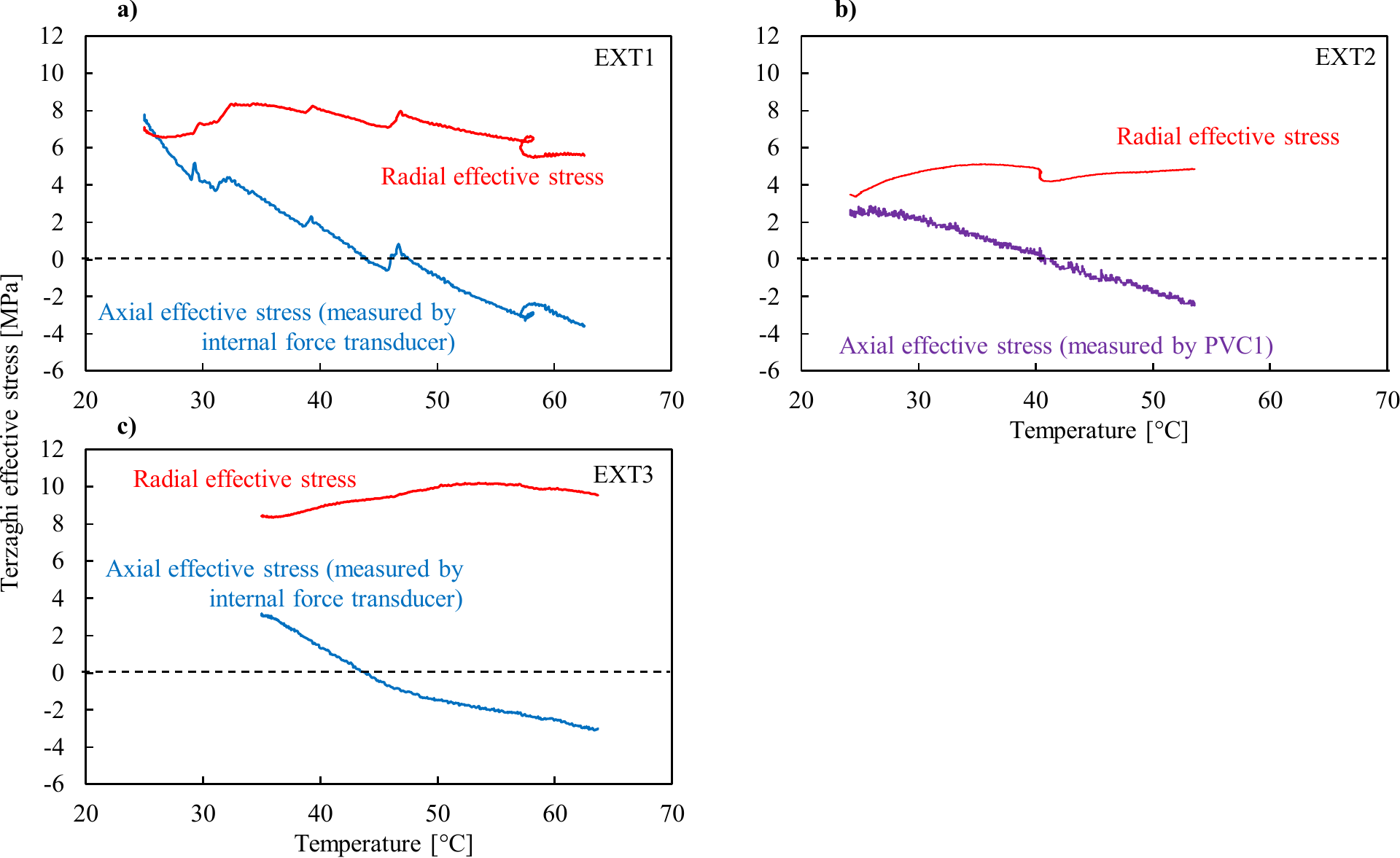}
\caption{Changes in Terzaghi effective stresses with temperature during the three heating tests, where negative stresses represent effective tension. In test EXT2 the internal force sensor malfunctioned and stresses measured by the external transducer \hl{are shown instead}.}
\label{fig:ext:stresses}      
\end{figure*}

Figure \ref{fig:ext:stresses} presents the changes in Terzaghi effective stress with increased temperature. The linear decrease of the effective axial stress is related to the linear increase in pore pressure observed in Figs. \ref{fig:ext1} to \ref{fig:ext3}. The temperatures at which axial effective tensile stress \hl{(negative effective stress)} starts to be applied are quite comparable in all three cases, at 43.5 $^{\circ}$C for EXT1 and EXT3 and 40.5 $^{\circ}$C for EXT2. Finally, as shown in Tab. \ref{tab:ext:results}, thermal tensile failure is observed at -3.6 MPa and 61.9 $^{\circ}$C (with a radial effective stress of 5.5 MPa) for EXT1, at -2.4 MPa and 53.4 $^{\circ}$C (with a radial effective stress of 4.8 MPa) for EXT2 and -3.0 MPa at 63.5 $^{\circ}$C (with a radial effective stress of 9.6 MPa) for EXT3. 
Interestingly, the changes in effective radial stress are rather small and limited between the boundary values (5.5 and 8.3 MPa for EXT1, 3.5 and 5.1 MPa for EXT2 and 8.5 and 10.2 MPa for EXT3). 

\begin{figure}
\includegraphics[width=0.5\textwidth]{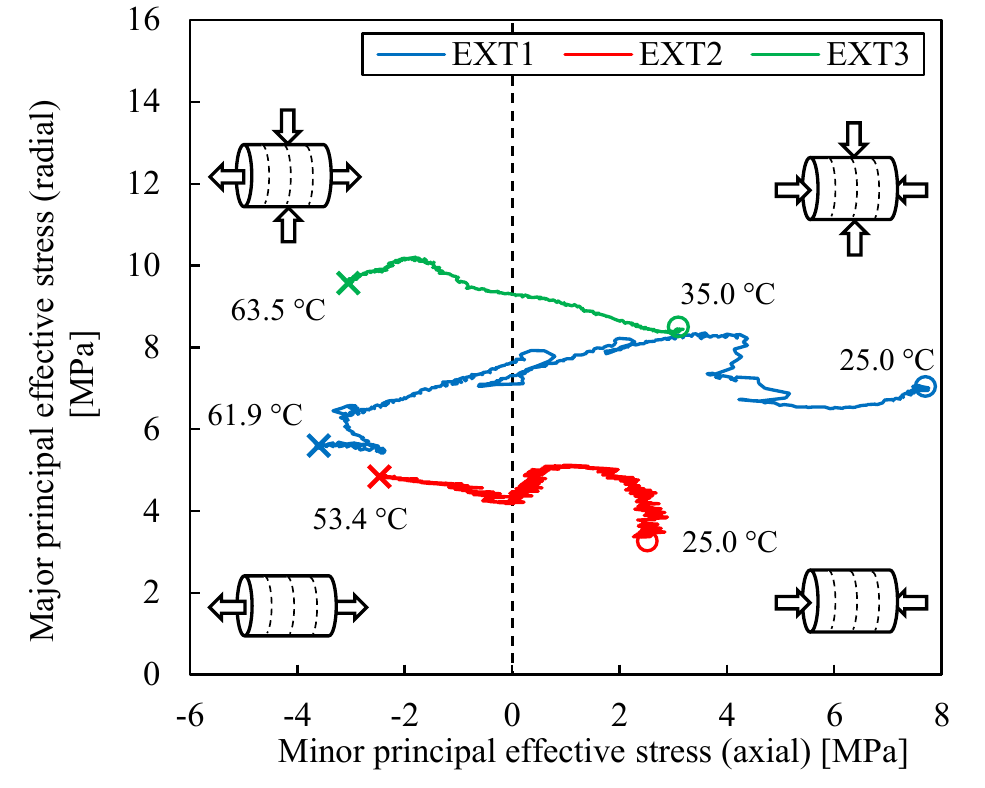}
\caption{Principal (Terzaghi) effective stress paths for all three specimens tested. Circles indicate the initial stress state and temperature, while crosses mark the specimen failure with their respective temperature.}
\label{fig:ext:failure}       
\end{figure}

The thermo-hydro-mechanical stress paths \hl{of} the three tests are represented in Fig. \ref{fig:ext:failure} in terms of radial ($y$-axis) and axial ($x$-axis) principal effective stress. 
The curves \hl{indicate} that there is no clear dependency of the thermal tensile failure stress with respect to the applied effective radial stress.

\hl{Note that the aim of this study is to reproduce stress paths similar to the ones in situ, in order to obtain a failure criterion close to the in situ conditions. Since laboratory and in situ conditions are hardly identical, undrained conditions are assumed in this study. These are neither exactly met in situ due to the complex 3D drainage conditions, nor in the laboratory due to the dead volume of the drainage system. 
In consequence, the failure criterion measured in this study has to be compared to the results of numerical simulations, which take into account complex 3D conditions of temperature, stress and drainage.}

\section{Thermo-poroelastic analysis}
\label{sec:ext:analysis}

Due to the complex nature of the transversely isotropic coupled THM behaviour, it is difficult to directly evaluate material parameters from the test data. 
\hl{In the presented complex loading paths, a variety of constitutive behaviours (elasticity, plasticity, viscosity, etc.) could be involved. For the following calculations, elastic behaviour is assumed. This simple analysis shall provide a basis for future in-depth modelling. In the following, a transversely isotropic THM model is used to simulate the conducted experiments. Material parameters are taken from different previous studies.}
The analysis is \hl{presented in two steps:
First, an element test is simulated in drained conditions, where the experimentally measured pore pressure is imposed. Only a single material point is simulated, therefore the system can be solved in an analytical way or using finite differences, if nonlinear parameters are considered. The simulated stress-strain response is compared with the experimental results, to calibrate thermo-poroelastic material properties.
If one wants to model the material behaviour in undrained conditions in a second step, the presence of the dead volume of the drainage system has to be taken into account. This dead volume induced pseudo-undrained conditions, which can be simulated based on previous calibration tests (Sec. \ref{sec:ext:calib}).} 


\subsection{Transversely isotropic thermo-porolastic framework}
\label{sec:ext:theory}

The thermal properties and THM couplings are analysed within the thermo-poroelastic constitutive equations, presented among others in the works of \cite{Biot195759}, \cite{Palciauskas198228} and \cite{Coussy2004}. The thermo-poromechanical formulations are written for a Representative Elementary Volume (REV) of the porous material.
A relationship between strain $\varepsilon_{i}$, total stress ${\sigma _i}$, pore pressure $p_{f}$ and temperature $T$ is expressed within this framework, as:
%
%
\begin{equation}
\label{eq:ext:epsi}
\mathrm{d}\varepsilon_{i}=C_{ij}  \left(   \mathrm{d}\sigma_{j} -  b_{j}\mathrm{d}p_f  \right)  - \alpha_{d,i}\mathrm{d}T
\end{equation}
where ${\sigma _j}$ is composed of the normal \hl{stresses} and shear \hl{stresses} in different directions:
\begin{equation}
\sigma_i = {\left[ {{\sigma_x},{\sigma_y},{\sigma_z},{\sigma_{xy}},{\sigma_{yz}},{\sigma_{zx}}} \right]^\top}
\end{equation}

The strains in different direction are contained within the strain vector $\varepsilon_i$:
\begin{equation}
{\varepsilon}_i = {\left[ {{\varepsilon_x},{\varepsilon_y},{\varepsilon_z},{\varepsilon_{xy}},{\varepsilon_{yz}},{\varepsilon_{zx}}} \right]^\top}
\end{equation}

In terms of material properties, \hl{one has} to define the Biot effective stress coefficients $b_i$, the compliance matrix $C_{ij}$ and the thermal expansion coefficients $\alpha_{d,i}$. 

It has been shown that the COx claystone is transversely isotropic \hl{\mbox{\citep{Plua2021}}}, \hl{therefore} a simplified set of coefficients \hl{is presented} (the properties along the directions $x$ and $y$ are identical, denoted in the following with a subscript $h$). Like in most claystones, this plane of isotropy corresponds to the bedding plane. 
According to \citet{Cheng199719}, the coefficients of transverse isotropy are recalled here.

The vector $b_i$ represents the Biot effective stress coefficents in both directions of anisotropy:
\begin{equation}
b_i = {\left[ {{b_h},{b_h},{b_z},0,0,0} \right]^\top}
\label{eq:ext:bi}
\end{equation}

\hl{One} can also write down the entries of the compliance matrix $C_{ij}$ as follows:
\begin{equation}
C_{ij} =  
\begingroup 
\setlength\arraycolsep{5pt}
\begin{pmatrix}
{{C_{11}}}&{{C_{12}}}&{{C_{13}}}&0&0&0\\
{{C_{12}}}&{{C_{11}}}&{{C_{13}}}&0&0&0\\
{{C_{13}}}&{{C_{13}}}&{{C_{33}}}&0&0&0\\
0&0&0&{1/2G'}&0&0\\
0&0&0&0&{1/2G}&0\\
0&0&0&0&0&{1/2G}
\end{pmatrix}
\endgroup
\end{equation}
with 
\begin{equation}
\label{eq:ext:entries_C}
\begin{array}{ll}
C_{11} &= 1/E_h	\\
C_{12} &= -\nu_{hh}/E_h	\\
C_{13} &= -\nu_{zh}/E_z	\\
C_{33} &= 1/E_z	\\
G'&=E_h/(1+\nu_{hh})	\\
\end{array}
\end{equation}
where ${E_z}$ and $\nu_{zh}$ are the Young modulus and the Poisson ratio perpendicular to the bedding plane, and ${E_h}$ and $\nu_{hh}$ parallel to the bedding plane, respectively. $G$ describes the independent shear modulus perpendicular to the isotropic plane.

The thermal behaviour is represented by $\alpha_{d,i}$, which consists of the two linear drained thermal expansion coefficients perpendicular ($\alpha_{d,z}$) and parallel ($\alpha_{d,h}$) to the bedding orientation: 
\begin{equation}
\label{eq:ext:advector}
\alpha_{d,i} = {\left[ {{\alpha_{d,h}},{\alpha_{d,h}},{\alpha_{d,z}},0,0,0} \right]^\top}
\end{equation}

The change of fluid content $m_f$ is described by \cite{Coussy2004}:
\begin{equation}
\label{eq:ext:fluid_mass_content}
\frac{{\rm{d}} m_f}{\rho_f}  =  - {b_i}{\rm{d}}{\varepsilon _i} + \frac{1}{M}{\rm{d}}p_f - \left( {b_i}{\alpha _{d,i}} - \phi {\alpha _\phi } +\phi\alpha_f  \right) {\rm{d}} T
\end{equation}
%
%
%
%

Here, $\alpha_{\phi}$ denotes the bulk thermal expansion coefficient of the pore volume and $\alpha_{f}$ the bulk thermal expansion coefficient of the pore fluid.
The parameter $M$ is the Biot modulus, which, acccording to \cite{Aichi201233}, can be calculated for a saturated transversely isotropic material:
\begin{equation}
\label{eq:ext:M}
\frac{1}{M} = 2(1 - {b_h})\left[ {\frac{{\left( {1 - {\nu _{hh}}} \right){b_h}}}{{{E_h}}} - \frac{{{\nu _{zh}}{b_z}}}{{{E_z}}}} \right] 
+ \frac{{\left( {1 - {b_z}} \right)}}{{{E_z}}}\left( {{b_z} - 2{\nu _{zh}}{b_h}} \right) + \phi \left( {\frac{1}{{{K_f}}} - \frac{1}{{{K_\phi }}}} \right)
\end{equation}
where ${K_f}$ the bulk modulus of the pore fluid and ${K_{\phi}}$ the bulk modulus of the pore volume.

\subsection{Simulation of a heating test \hl{with imposed pore pressure}}
\label{sec:ext:strain_calc}

First, the observed claystone behaviour \hl{is simulated during} extension using the thermo-poroelastic framework discussed in Sec. \ref{sec:ext:theory}. \hl{Element tests are modelled with homogeneous pore pressure, temperature, stress and strain distributions. Linear thermo-poroelasticity is considered, therefore the experiments can be reproduced through analytical calculations.} 
Zero lateral deformation, the measured axial stress, the recorded pore pressure changes and the applied temperature \hl{are imposed}. 
Constitutive parameters were taken from the literature, summarized in \hl{the following:} 
\cite{Escoffier2002} and \cite{Belmokhtar201787} provided measurements for $b$ close to 0.9 at effective stresses close to the in situ one (around 9 MPa). Recently, \cite{Braun2020el} estimated the Biot coefficient also for lower effective \hl{stress} levels, showing that $b_z$ approaches 1.0 at effective stresses tending to zero. For \hl{the following} calculations, representing the in situ condition where the mean effective stress decreases, starting from around 8 MPa, $b_z=b_h=0.9$ \hl{was chosen}. The Young modulus $E_z$ at 8 MPa mean effective stress was determined by \cite{Menaceur201529} and \cite{Belmokhtar201819}, with values around 3 GPa, and at lower mean effective stress by \cite{Menaceur201529} and \cite{Zhang201279}, with values around 1.0 and 1.5 GPa. For 8 MPa effective mean stress, \cite{Braun2020el} found through a multivariate regression a best-fit value of $E_z$ close to 2.6 GPa, which was adopted in this calculation. 
Less data is available for $E_h$ and the Poisson ratios. $E_h=$ 5.7 GPa and $\nu_{zh}=0.11$, which was measured by  \cite{Braun2020el} under 8 MPa effective stress \hl{is utilized}. $\nu_{hh}=0.29$ was given by the Andra database \citep{Guayacan-Carrillo201701} and confirmed by \cite{Braun2020el}. 
\hl{The anisotropic drained thermal expansion coefficients $\alpha_{d,z} = 0.21 \times 10^{-5}{^\circ \mathrm{C}}^{-1}$ and $\alpha_{d,h}=0.51 \times 10^{-5}{^\circ \mathrm{C}}^{-1} $ are taken from the experimental results of \mbox{\cite{Braun2020th}} on COx claystone. Moreover, a porosity of 0.18, measured in the present study, is used.}   
\hl{It is assumed} that all parameters are stress and temperature independent.

\begin{figure*}
\centering
  \includegraphics[width=\textwidth]{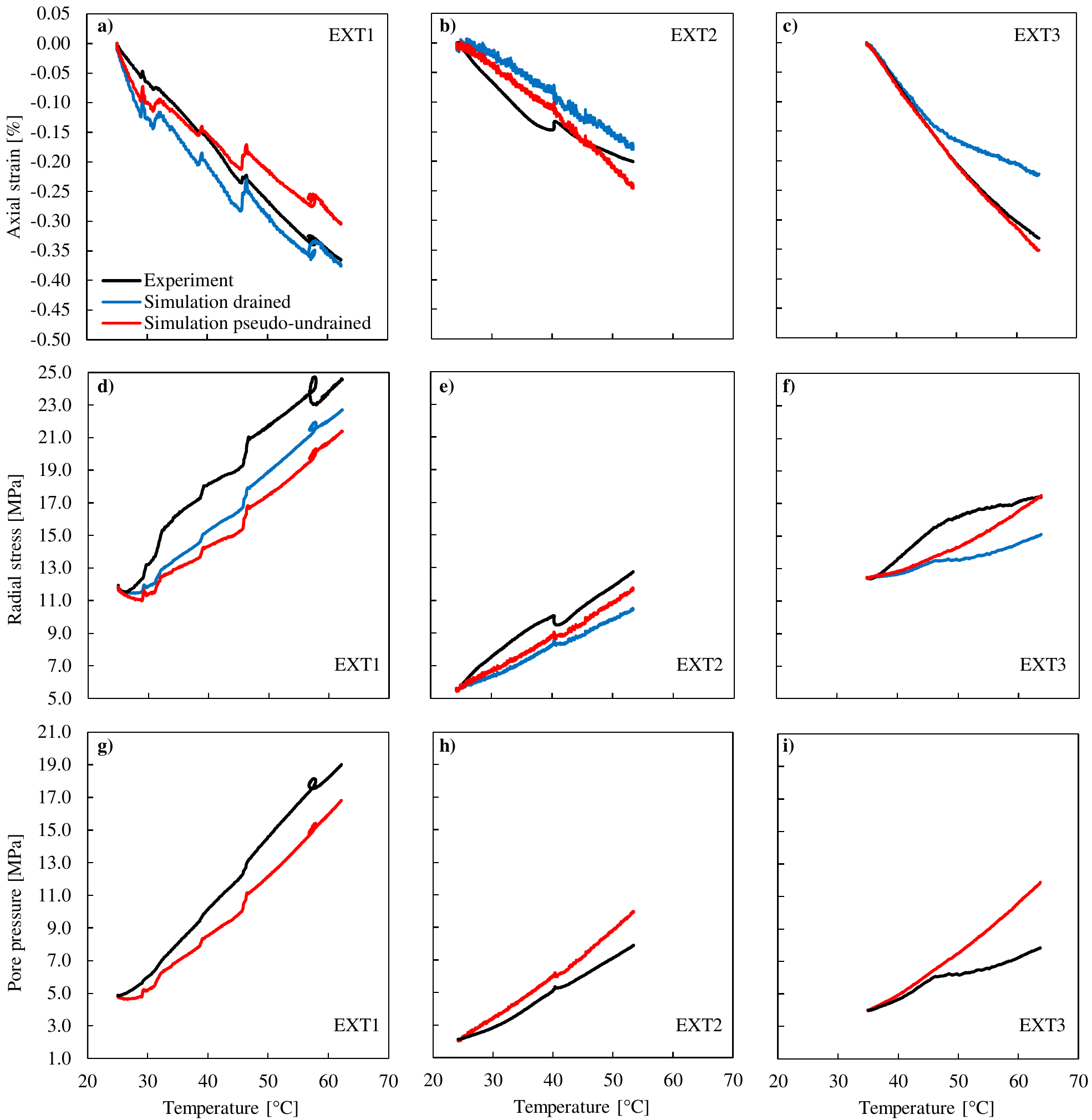}
\caption{Experimental results and calculated behaviour using a thermo-poroelastic model on three extension tests EXT1-3. a) - c) Temperature - axial strain evolution, d) - f) radial stress evolution with temperature and g) - i) thermally induced pore pressure increase. Note that for the simulation in drained conditions, the pore pressure increase from experimental data was imposed.}
\label{fig:ext:simulation}     
\end{figure*}

Using Eq. (\ref{eq:ext:epsi}), one is able to calculate the changes in radial stress and axial strains, based on the applied axial stress, the measured pore pressure and the applied temperature (Fig. \ref{fig:ext:simulation}a-f). The radial strains were kept equal to zero. The strain calculations show a good match with the experimental data, without any parameter fitting necessary. \hl{Only} the axial strains in EXT3 become significantly nonlinear at around 50 $^{\circ}$C, which is not represented by the simulation. This difference could be due to natural variability of the specimens, the occurrence of strain localisation, not detected  local strain gages, or induced damage or plastic deformations, which cannot be captured with the linear thermo-poroelastic model. 
Calculated radial stresses show a general slight underestimation. \hl{To get a first estimation of the parameter sensitivity, a preliminary parametric analysis was carried out within the following section.} 

\subsection{Simulation of thermal pressurization in pseudo-undrained conditions}
\label{sec:ext:press_calc}

To simulate the experiment in pseudo-undrained conditions, the dead volume of the drainage system \hl{can be modelled} through Eq. (\ref{eq:ext:drainage_balance}), \hl{using} the drainage system properties evaluated in Sec. \ref{sec:ext:calib}.
The change of the specimen fluid mass per unit volume is calculated through Eq. (\ref{eq:ext:fluid_mass_content}).
A single element test with uniform pore pressure and fluid density within the specimen and the drainage system ($p_f=p_L$, $\rho_f=\rho_L$) \hl{is simulated}. 
Next to the properties used in Sec. \ref{sec:ext:strain_calc}, additional parameters concerning the deformations of the pore space and the pore fluid are required in Eq. (\ref{eq:ext:fluid_mass_content}). The Biot modulus is computed through Eq. (\ref{eq:ext:M}). Moreover, as a first hypothesis, \hl{it is} assumed that $K_\phi$ \hl{is equal to the unjacketed bulk modulus of} 19.7 GPa \citep{Braun2020el} and $\alpha_\phi$ \hl{is equal to} $2\alpha_h+\alpha_z$.
For the water properties $\alpha_f$ and $K_f$, \hl{a dependency on} the fluid pressure and temperature \hl{is considered, according to} \cite{IAPWS-IF972008}. \hl{In consequence, the material response becomes nonlinear, which was simulated using the explicit finite difference method} with sufficiently small increments ${\rm{d}}T \leq$ 0.1 $^{\circ}$C, ${\rm{d}}p_f \leq$ 0.2 MPa.

\hl{One observes} a good simulation of the axial extension strains in Fig. \ref{fig:ext:simulation}a-c, the radial stress change in Fig. \ref{fig:ext:simulation}d-f and the thermally induced pore pressures in Fig. \ref{fig:ext:simulation}g-i. Axial strain changes are well reproduced by the simulation, while radial stresses of EXT1 and EXT2 are slightly underestimated. Calculated pore pressures follow the experimental trend, while they slightly overestimate the result of test EXT1 and underestimate EXT2. The pore pressure increase of EXT3 is significantly overestimated.

Due to the parallel over- and underestimation of experimental pore pressures, an adjustment of the additional parameters $K_\phi$, $\alpha_\phi$, $\alpha_f$ and $K_f$ does not improve the fitting. \hl{A preliminary parametric analysis was carried out by varying separately each model parameter by 10 \% and observing the variation of the final pore pressure in test EXT1. The highest parameter sensitivity was detected for $b_z$, which caused a variation of the resulting pore pressure of 7 \%. The parameters $\alpha_f$ and $\phi$ caused a variation of 4 \% and $E_z$ a variation of 3 \%. The variation of other parameters induced a change of $\leq 1\%$ on the pore pressure result.
Also a better fitting of the horizontal elastic properties $E_h$ and $\nu_{zh}$, on which literature data is scarce, could provide better results. For further information on the in situ variability of parameters, the reader is referred to \mbox{\cite{Plua2021}}.}

The set of linear thermo-poroelastic claystone properties used here appears to be sufficient to simulate closely the measure laboratory thermal extension. It has to be noted that non-linear water properties are essential to model thermal pressurization.


%

\section{Failure criterion}
\label{sec:ext:fail}  

\begin{figure}
  \includegraphics[width=0.5\textwidth]{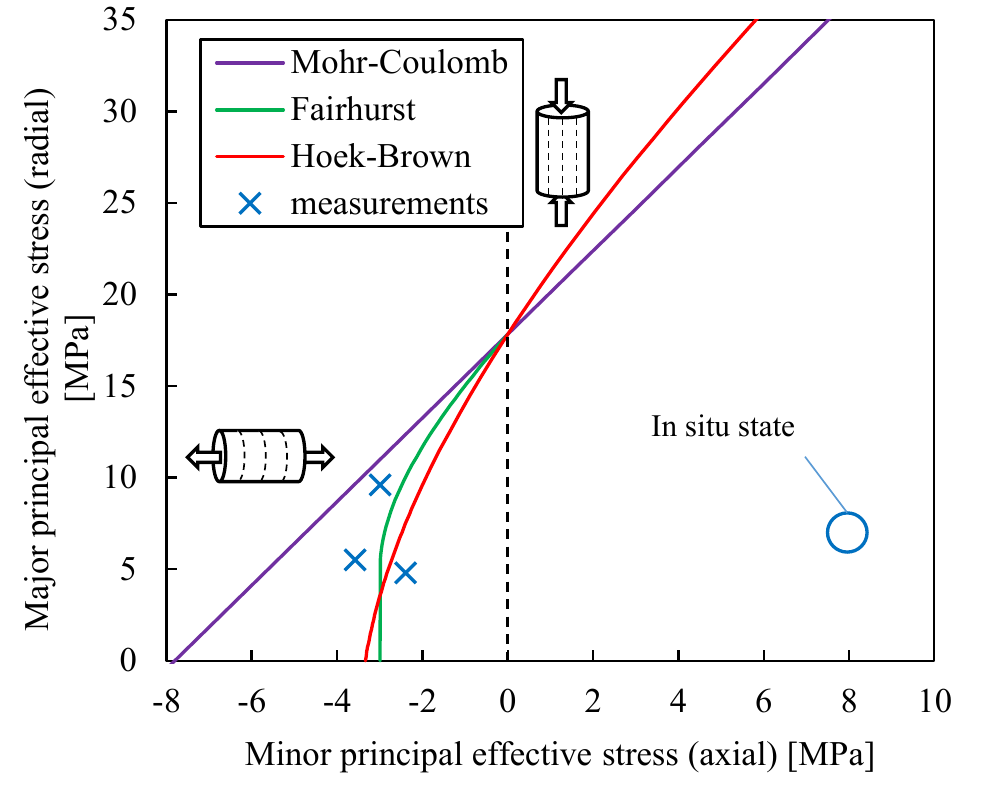}
\caption{Hoek-Brown and Fairhurst's generalized Griffith criterion, adopted for measured tension failure with tension perpendicular to the bedding. The Mohr-Coulomb criterion is obtained from the Andra database, based on shear failure with a major principal stress parallel to bedding}
\label{fig:ext:2failurecriterion}      
\end{figure}

Based on the stress state measured at tensile failure (Fig. \ref{fig:ext:failure}), the existing shear failure criterion of the COx claystone \hl{is extended}.
A Mohr-Coulomb (MC) type failure criterion, based on uniaxial and triaxial compression tests on the COx claystone, is given in the Andra database (applied also by \citealp{Guayacan-Carrillo201701}). The criterion can be transformed to the principal effective stress domain, written as:
\begin{equation}
\label{eq:ext:MC}
{\sigma _1'} = \frac{{1 + \sin \varphi }}{{1 - \sin \varphi }}{\sigma _3'} + \frac{{2c\cos \varphi }}{{1 - \sin \varphi }}
\end{equation}  

\noindent where ${\sigma _1'}$ and ${\sigma _3'}$ are the major and minor principal effective stresses, respectively. Values for the cohesion $c=$ 5.9 MPa and the friction angle $\varphi=$ 23$^{\circ}$ are taken from the Andra database, when the minor effective stress is perpendicular to the bedding plane. This shear failure criterion, displayed in Fig. \ref{fig:ext:2failurecriterion}, provides us an unconfined compressive strength ${\sigma _{c}}={\sigma _1'}({\sigma _3'}=0)=17.8$ MPa. 
One notes, that the Mohr-Coulomb criterion overestimates the measured strength in tension, and is therefore not applicable in this region.

\cite{Hoek198010} and \cite{Hoek198318} described the Hoek-Brown \hl{(HB)} failure criterion, proposed as an empirical relationship for observed shear failure of rock under triaxial compressive stress:

\begin{equation}
\label{eq:ext:HB}
\sigma {'_1} = \sigma {'_3} + {\sigma _c}{\left( {{m_i}\frac{{\sigma {'_3}}}{{{\sigma _c}}} + 1} \right)^{0.5}}
\end{equation}  

This criterion \hl{was fitted} for the measured strength in tension. The shear failure region was not investigated here, however ${\sigma _{c}}=17.8$ MPa from the Andra database \hl{can be adopted}. A single parameter $m_i$ had to be fitted, carried out by minimising the sum of squared errors in terms of $\sigma'_3$. The best-fit, obtained with $m_i=$ 5.15, is presented in Fig. \ref{fig:ext:2failurecriterion}.  
\hl{Note that the HB criterion was conceived for intact isotropic rock, while the COx claystone is anisotropic \mbox{\citep{Eberhardt2012}}. For a more versatile form of the HB criterion, the reader is referred to the "generalised" HB criterion \mbox{\citep{Hoek1995}}.} 

As quoted by \cite{Hoek201428}, the HB criterion can over-estimate the tensile strength of rocks with a low $m_i$ value. \cite{Hoek201428} suggest to utilize the Fairhurst generalized Griffith (FG) failure criterion \citep{Fairhurst196453} to predict tensile failure. This criterion can be applied in the form of a so-called "tension cutoff" under tensile stresses, combined with a Mohr-Coulomb or Hoek-Brown criterion for shear stresses.

\cite{Fairhurst196453} presented the FG criterion \citep{Griffith192455}, which is here recalled according to \cite{Hoek201428}:

If $w(w-2) \sigma_3' +\sigma_1' \leq 0$, failure occurs when:
\begin{equation}
\label{eq:ext:straingeneral}
{{\sigma _3'} = {\sigma _t}}
\end{equation}  
or when:
\begin{equation}
\label{eq:ext:straingeneral}
{\sigma _1'} = \frac{\left( {2{\sigma _3'} - A{\sigma _t}} \right)}{2}
 +\frac{ \sqrt {{{\left( {A{\sigma _t} - 2{\sigma _3'}} \right)}^2} - 4\left( {\sigma _3'^2 + A{\sigma _t}{\sigma _3'} + 2AB\sigma _t^2} \right)} }{2}
\end{equation}  
where
\begin{equation}
\label{eq:ext:straingeneral}
\left. {\begin{array}{*{20}{l}}
{A = 2{{\left( {w - 1} \right)}^2}}\\
{B = {{\left[ {\left( {w - 1} \right)/2} \right]}^2} - 1}\\
{w = \sqrt {{\sigma _c}/\left| {{\sigma _t}} \right| + 1} }
\end{array}} \right\}
\end{equation}  

Combined with the Mohr-Coulomb criterion described before, the major principal stress where both criteria coincide at ${\sigma _{3}'=0}$, is the unconfined compressive strength ${\sigma _{c}}=17.8$ MPa. The remaining coefficient to be determined for the FG criterion is the tensile strength ${\sigma _{t}}$, that can be evaluated from the laboratory measurements. A regression \hl{by} minimizing the orthogonal least square error \hl{was used}, which provided the best-fit value of ${\sigma _{t}=}$ 3.0 MPa (Fig. \ref{fig:ext:2failurecriterion}). 

Both HB and FG criteria are able to satisfactorily characterize the tensile failure of the COx specimens. Both criteria can be adopted in combination with other criteria for shear failure, such as the MC criterion presented before. The HB criterion fitted for tensile failure was observed to overestimate the shear failure characterized by a MC criterion from the Andra database. 

\hl{Note that this data was fitted for failure under tension perpendicular to the bedding layer. Both HB and FG criteria account only for isotropic resistance. To consider anisotropic resistance, one would have to modify the criterion by adjusting $m_i$, ${\sigma _{t}}$ or ${\sigma _{c}}$ with respect to the principal loading direction (see also \mbox{\citealp{Pardoen2015,Manica2021}}).}

The characteristics evaluated in the thermal extension tests of this study are compared with literature data on other shales \hl{and claystones} in Tab. \ref{tab:ext:criterion_literature}.
One observes that the parameter $m_i$ ($\sigma_t=\sigma_c/m_i$) for the COx claystone is found within the range of $m_i \approx$ 4.0 \citep{Hoek200719} and within the estimated values for other shales \hl{and claystones}. \hl{One notes} that the estimated values of $m_i$ are all somewhat higher than 4.0. The measured tensile resistance of the COx claystone is comparable to that of other shales \hl{and claystones}, in the range of a few MPa. 

The values obtained here by direct tension tests are somewhat higher than those measured by \cite{Auvray201549} indirectly through Brazilian tests. This could be due to difference in testing methods. The samples of \cite{Auvray201549} were saturated under stress, but had to be de-confined for testing, which could result in some damage. Note also, that probably due to their small size, the specimens tested in the present study were prone to breakage during sample preparation. This could have possibly resulted an involuntarily rejection of less resistant or pre-fractured specimens, so that predominantly resistant samples were finally tested.   

\cite{Auvray201549} did not note any significant change of tensile resistance on specimens tested at 90 $^{\circ}$C, neither could a temperature dependency be observed in this study, due to the scarcity of results.

\section{Conclusions}
\label{sec:ext:conclusion}

A novel triaxial apparatus for investigating the thermal pressurization failure of soft rocks was developed and \hl{used on} samples of the COx claystone. The challenging conditions, \hl{which might} occur at mid-distance between two horizontal and parallel micro-tunnels containing high activity exothermic radioactive wastes, are mimicked by the designed device.
At this location lateral strains are constrained by symmetry, the overburden stress is constant \hl{and undrained heating by the two adjacent micro-tunnels increases the temperature. These loadings} submit the claystone to an extension, which could \hl{eventually} lead to tensile failure. 

A triaxial device with simultaneous control of temperature, axial and radial stresses and pore pressure was designed and employed. \hl{The lateral strain was kept to zero} by servo-controlling the confining stress, while maintaining the axial stress \hl{constant} through the piston. 
\hl{Axial effective tensile stresses require gluing the top and bottom caps through epoxy resin with a higher tensile strength than the tested specimen. Special bases with screw connections were designed for that purpose. For specimen saturation, it was necessary to implement special drainage ducts and a lateral geotextile. Axial stress was measured internally by a calibrated aluminium force transducer. 
To achieve precise real-time lateral strain control, measured strains had to be corrected for temperature effects through a dummy strain gage. Attention had to be paid to tuning the proportional parameter of the PID controller, which determines the confining stress response upon radial stain changes. The derivative parameter was not necessary for fast control. 
}

\hl{The investigated cylindrical COx specimens of 20 mm diameter and 30 – 40 mm height, cored perpendicular to the bedding plane, were handled with care to avoid drying or swelling. All specimens were saturated under close to in situ effective stress.} A series of undrained heating tests with zero radial strain and constant axial stress was carried out on these samples. 
As expected, an increase in pore pressure due to thermal pressurization was observed. Constraining the radial strain to zero required a significant increase in radial total stress. \hl{Pore pressure and radial stress increased for nearly the same amount, leading to a fairly constant radial effective stress. Axial effective stress decreased continuously until reaching negative values. The recorded axial strains showed a fairly linear extension behaviour.}
Temperature induced pressurization led to failure, once the limit effective tensile strength of the claystone was reached at values around 3 MPa and temperatures between 53 and 64°C.

\hl{The observed strength of COx claystone under tension perpendicular to the bedding plane is overestimated by criteria calibrated on triaxial compression tests. A Hoek-Brown and a Fairhurst generalized Griffith criterion, calibrated on the rock’s measured tensile strength, are proposed for a better representation of the failure in the low effective stress regime. The tensile strength parallel to the bedding plane was not measured in this study and might be different to the perpendicular one. Anisotropic resistance could be taken into account by failure criterion parameters which depend on the stress orientation.
The size effect on the tensile resistance was not investigated in this study. According to Ba{\v{z}}ant's theory \mbox{\citep{Bazant1984}}, the tensile resistance of larger samples might be smaller than the one measured in this study. Also there might be a higher probability for the existence of heterogeneities in larger samples. Future experiments on samples with larger diameter could provide important insight in this regard.} 

Satisfactory modelling of the experimental response was achieved within a thermo-poroelastic framework, based on thermo-hydro-mechanical parameters of the COx claystone that were previously determined by \cite{Braun2020th,Braun2020el}. \hl{Axial strains were well reproduced by the thermo-poroelastic simulation. Only some non-linear behaviour of experiment EXT3 could not be simulated. One could observe that calculated radial stresses of two tests show a slight underestimation of measured behaviour. The simulated pore pressure changes follow in general the experimental trend with no clear over- or underestimation. The pore pressure evolution was found to be most sensitive to the Biot coefficient $b_z$, followed by $\alpha_f$, $\phi$ and $E_z$. Note that the adoption of non-linear water properties is essential to model thermal pressurization.}

\section*{Conflict of interest}

The authors declare that there are no known conflicts of interest associated with this publication.

\bibliographystyle{spbasic}      
\bibliography{manuscript}   

%
%

\end{document}